\documentclass[usenatbib]{mnras}
\usepackage{natbib,amssymb,amsmath,times,graphicx,url,aas_macros}
\usepackage{setspace}      
\usepackage{epstopdf}       
\usepackage{verbatim}       
\usepackage{bm}		  
\usepackage{fleqn}		  
\usepackage{Wurster_macros}

\newcommand\bmx{$\bm{B}_\text{-x}$}
\newcommand\bmz{$\bm{B}_\text{-z}$}
\newcommand\bpz{$\bm{B}_\text{+z}$}
\newcommand\bpmz{$\bm{B}_{\pm\text{z}}$}
\newcommand\bbmx{$\bm{B}_0 = \bm{B}_\text{-x}$}
\newcommand\bbmz{$\bm{B}_0 = \bm{B}_\text{-z}$}

\newcommand\bbpmz{$\bm{B}_0 = \bm{B}_{\pm \text{z}}$}
\newcommand\bbpmx{$\bm{B}_0 = \bm{B}_{\pm \text{x}}$}

\newcommand\omegaz{$\Omega_0$}
\newcommand\omegaeq[1]{$\Omega_0 = #1\Omega_\text{orbit}$}

\newcommand\omegage[1]{$\Omega_0 \ge #1\Omega_\text{orbit}$}

\newcommand\mmodel[4]{{#4}$\Omega_{#1}\mu_{#2}B_\text{#3}$}
\newcommand\imodel[3]{I$\Omega_{#1}\mu_{#2}B_\text{#3}$}
\newcommand\nmodel[3]{N$\Omega_{#1}\mu_{#2}B_\text{#3}$}
\newcommand\hmodel[1]{H$\Omega_{#1}$}

\title[Disc formation and fragmentation]{Disc formation and fragmentation using radiative non-ideal magnetohydrodynamics}

\author[Wurster \& Bate]{James Wurster$^{1}$\thanks{j.wurster@exeter.ac.uk} and Matthew R. Bate$^{1}$\thanks{mbate@astro.ex.ac.uk}  \\
$^{1}$School of Physics and Astronomy, University of Exeter, Stocker Rd, Exeter EX4 4QL, UK \\
}
\date{Submitted: Revised: Accepted: }
\pagerange{\pageref{firstpage}--\pageref{lastpage}} \pubyear{2016}

\begin{document}
\label{firstpage}
\bibliographystyle{mnras}
\maketitle

\begin{abstract}
We investigate the formation and fragmentation of discs using a suite of three-dimensional smoothed particle radiative magnetohydrodynamics simulations.  Our models are initialised as 1M$_\odot$ rotating Bonnor-Ebert spheres that are threaded with a uniform magnetic field.  We examine the effect of including ideal and non-ideal magnetic fields, the orientation and strength of the magnetic field, and the initial rotational rate.  We follow the gravitational collapse and early evolution of each system until the final classification of the protostellar disc can be determined.   Of our 105 models, 41 fragment, 21 form a spiral structure but do not fragment, and another 12 form smooth discs.  Fragmentation is more likely to occur for faster initial rotation rates and weaker magnetic fields.  For stronger magnetic field strengths, the inclusion of non-ideal MHD promotes disc formation, and several of these models fragment, whereas their ideal MHD counterparts do not.  For the models that fragment, there is no correlation between our parameters and where or when the fragmentation occurs.  Bipolar outflows are launched in only 17 models, and these models have strong magnetic fields that are initially parallel to the rotation axis.  Counter-rotating envelopes form in four slowly-rotating, strong-field models -- including one ideal MHD model -- indicating they form only in a small fraction of the parameter space investigated.
\end{abstract}


\section{Introduction}
\label{intro}

Stars are formed from the gravitational collapse of gas clouds, which are observed to have strong magnetic fields \citep[e.g.][]{HeilesCrutcher2005} and low ionisation rates \citep[e.g.][]{MestelSpitzer1956,NakanoUmebayashi1986,UmebayashiNakano1990}.  Dense molecular cloud cores within these clouds are initially rotating with rates of $\beta_\text{r} \lesssim 0.15$ with a mean value of $\beta_\text{r} \sim 0.02$ \citep{Goodman+1993}, where $\beta_\text{r}$ is the ratio of rotational to gravitational energy.  Since the dense core is initially rotating, a self-gravitating disc is expected to  form during the formation of the protostar \citep[e.g.][]{TerebeyShuCassen1984,Attwood+2009,Bate2011,MachidaMatsumoto2011}, and large gas discs have been inferred to exist around even young (Class 0) protostars \citep[e.g.][]{Dunham+2011,Lindberg+2014,Tobin+2015,Tobin+2016triple}.

During the gravitational collapse, a massive first hydrostatic core can become bar unstable to form a gravitationally unstable disc and develop spiral arms \citepeg{Bate1998,SaigoTomisaka2006,SaigoTomisakaMatsumoto2008}.  Alternatively, if a self-gravitating disc forms, it may undergo a gravitational instability due to growing non-axisymmetric modes and then form spiral arms \citepeg{PapaloizouSavonije1991,Durisen+2007}.  The disc itself, or more typically the spiral arms, may further be unstable to fragmentation \citepeg{Bonnell1994,BonnellBate1994md}.  There are several criteria that are used to predict the stability of a disc, including the Toomre-Q parameter \citep{Toomre1964}, the cooling parameter \citep{Gammie2001}, and the ratio of disc-to-stellar masses \citep{Gammie2001}. 

The study of disc fragmentation is important for better understanding the formation of multiple systems \citep[e.g.][]{KratterMatzner2006,NayakshinCuadraSpringel2007,Kratter+2010}, the formation of brown dwarfs or planets \citep[e.g.][]{Boss1997,Boss1998,Boss2001,MayerQuinnWadsleyStadel2002,StamatellosHubberWhitworth2007}, and episodic accretion events \citep[e.g.][]{VorobyovBasu2005,VorobyovBasu2006,VorobyovBasu2015}.  These studies have been performed starting from both gas clouds \citep[e.g.][]{StamatellosWhitworthHubber2011,ForganRice2012}, and from Keplerian discs \citep[e.g.][]{MayerQuinnWadsleyStadel2002,Rice+2003,StamatellosWhitworth2009,MeruBate2010,MeruBate2012,Vorobyov2013,Meru2015,ForganPriceBonnell2017,MercerStamatellos2017,HallForganRice2017}.   While many studies stop once the fragmentation limit is determined from their initial conditions and physical processes, several studies \citep[e.g.][]{VorobyovBasu2015,HallForganRice2017,ForganPriceBonnell2017} continue to evolve the system to further study the disc and the fragments.  

Studies starting from a Keplerian disc typically neglect magnetic fields since discs are expected to be weakly ionised (i.e. with an ionisation fraction of $10^{-12}$;  \citealp{FromangTerquemBalbus2002} and references therein), and contain magnetic dead zones in which the magnetic Reynolds number is lower than some critical value \citep[e.g.][]{Gammie1996}.  However,  \citet{ForganPriceBonnell2017} modelled a disc in the presence of ideal magnetic fields and found that the fragments in the magnetised disc were larger and formed at a smaller range of radii than in their purely hydrodynamical counterpart, and that these properties were dependent on the initial magnetic field strength.  They concluded that magnetic fields influence the fragmentation limit.

Studies that start from a gas cloud and include magnetic fields to match observations have shown that the resulting disc properties are at least partly dependent on the initial magnetic field strength and geometry \citep[e.g.][]{AllenLiShu2003,PriceBate2007,HennebelleFromang2008,DuffinPudritz2009,HennebelleCiardi2009,Commercon+2010,Seifried+2011,LewisBatePrice2015}.  These early studies starting from a gas cloud used ideal magnetohydrodynamics (MHD), and suffered from the `magnetic braking catastrophe' \citep{AllenLiShu2003} since the magnetic fields so efficiently extracted angular momentum that large discs failed to form in the presence of strong magnetic fields.

Detailed models of molecular clouds found have ionisation fractions as low as $10^{-14}$ \citep{NakanoUmebayashi1986,UmebayashiNakano1990}, suggesting ideal MHD is a poor approximation.  Rather than neglecting magnetic fields and incorrectly assuming a purely hydrodynamic collapse, simulations began including non-ideal MHD to account for ionised and neutral species \citep[e.g.][]{CiolekMouschovias1994,LiShu1996,MellonLi2009,DuffinPudritz2009,DappBasu2010,MachidaInutsukaMatsumoto2011,LiKrasnopolskyShang2011,Tomida+2013,TomidaOkuzumiMachida2015,Tsukamoto+2015hall,Tsukamoto+2015oa,WursterPriceBate2016,Tsukamoto+2017,WursterBatePrice2018sd,WursterBatePrice2018hd}.  The more recent simulations that included the Hall effect were able to overcome the magnetic braking catastrophe and produce discs comparable with observed discs.  These non-ideal MHD studies typically focused on disc formation rather than fragmentation, thus there has yet to be a fragmentation study using non-ideal MHD starting from a molecular cloud core.

In this study, we model the formation and fragmentation of discs using a 3D self-gravitating, smoothed particle, radiative, non-ideal magnetohydrodynamics code.  We self-consistently form a disc by allowing a low-mass molecular cloud to collapse, and we follow the evolution until a disc forms and it dissipates, is proved to be stable, or it fragments.  Thus, we primarily aim to determine the initial conditions that will lead to a fragmenting disc; we do not follow the evolution of the fragments.  In Section~\ref{sec:methods} we describe our methods and in Section~\ref{sec:ic} we give our initial conditions.  Results are presented and discussed in Sections~\ref{sec:results} and \ref{sec:discussion}, respectively; given the large suite, the discussion also includes aspects that are not directly related to fragmentation, but provide additional insights into the effect of varying our free parameters.   We conclude in Section~\ref{sec:conclusion}.

\section{Methods}
\label{sec:methods}
\subsection{Radiative non-ideal magnetohydrodynamics}

We solve the set of radiative non-ideal MHD equations, given by
\begin{eqnarray}
\frac{{\rm d}\rho}{{\rm d}t} & = & -\rho \nabla\cdot \bm{v}, \label{eq:cty} \\
\frac{{\rm d} \bm{v}}{\rm{d} t} & = & -\frac{1}{\rho}\bm{\nabla} \left[\left(P+\frac{1}{2}B^2\right)\mathbb{I} - \bm{B}\bm{B}\right] \notag \\ &-& \nabla\Phi + \frac{\kappa\bm{F}}{c}, \label{eq:mom} \\
\frac{{\rm d}}{\text{d} t}\left(\frac{\bm{B}}{\rho}\right) & = & \left(\frac{\bm{B}}{\rho}\cdot\bm{\nabla}\right)\bm{v}+ \frac{1}{\rho}\left.\frac{\text{d} \bm{B}}{\text{d} t}\right|_\text{non-ideal} \label{eq:ind}, \\
\rho \frac{{\rm d}}{\text{d} t}\left(\frac{E}{\rho}\right) & = & -\nabla\cdot\bm{F} - \nabla\bm{v}:\bm{P} + 4\pi\kappa\rho B_\text{P} - c\kappa\rho E \label{eq:rad},\\
\rho \frac{{\rm d} u}{\text{d} t} & = & -p\nabla\cdot{v} -  4\pi\kappa\rho B_\text{P} + c\kappa\rho E + \rho \left.\frac{\text{d} u}{\text{d} t}\right|_\text{non-ideal} \label{eq:u}, \\
\nabla^{2}\Phi & = & 4\pi G\rho, \label{eq:grav}
\end{eqnarray}
where $\text{d}/\text{d}t \equiv \partial/\partial t + \bm{v}\cdot\bm{\nabla}$ is the Lagrangian derivative,  $\rho$ is the density, $\bm{v}$ is the velocity, $P$ the hydrodynamic pressure, $\bm{B}$ is the magnetic field, $\Phi$ is the gravitational potential, $B_\text{P}$ is the frequency-integrated Plank function, $E$ is the radiation energy density, $\bm{F}$ is the radiative flux, $\bm{P}$ is the radiation pressure tensor, $\kappa$ is the opacity, $u$ is the specific energy of the gas, $\mathbb{I}$ is the identity matrix, $c$ is the speed of light and $G$ is the gravitational constant; the magnetic field has been normalised such that the Alfv{\'e}n velocity is defined as $v_\text{A}\equiv B/\sqrt{\rho}$ in code units \citep[see][]{PriceMonaghan2004}.   The radiative transfer algorithm is given in \citet{WhitehouseBateMonaghan2005} and \citet{WhitehouseBate2006} and uses a two-temperature (matter and radiation) flux-limited diffusion approximation and assumes local thermodynamic equilibrium; the opacity is assumed to be independent of frequency and there is no distinction between absorption and total opacities.

The contribution of the non-ideal MHD processes to the magnetic field and the internal energy are
\begin{flalign}
\label{eq:ni:B}
\left.\frac{\text{d} \bm{B}}{\text{d} t}\right|_\text{non-ideal} = &-\bm{\nabla} \times  \left[ \eta_\text{OR} \bm{J} +   \eta_\text{HE} \bm{J} \times\bm{\hat{B}} -  \eta_\text{AD}\left(\bm{J} \times\bm{\hat{B}}\right)\times\bm{\hat{B}}\right], &
\end{flalign}
and
\begin{flalign}
\label{eq:ni:u}
\left.\frac{\text{d} u}{\text{d} t}\right|_\text{non-ideal} &= \frac{\eta_\text{OR}}{\rho}\left|\bm{J}\right|^2 + \frac{\eta_\text{AD}}{\rho}\left[ \left|\bm{J}\right|^2 - \left(\bm{J}\cdot\hat{\bm{B}}\right)^2 \right], &
\end{flalign}
respectively, where $\bm{J} = \bm{\nabla} \times \bm{B}$ is the current density, and $\eta_\text{OR}$, $\eta_\text{HE}$ and $\eta_\text{AD}$ are the non-ideal MHD coefficients for Ohmic resistivity, the Hall effect and ambipolar diffusion, respectively.  

We use version 1.2.1 of the \textsc{Nicil} library \citep{Wurster2016} to calculate the non-ideal coefficients, $\eta$.  At low temperatures, cosmic rays ionise a heavy ion and a light ion at the canonical rate of $\zeta_\text{cr} = 10^{-17}$ s$^{-1}$ \citep{SpitzerTomasko1968,UmebayashiNakano1981}.  A single dust grain population is modelled as three species: a positively (negatively) charged grain species that has lost (absorbed) an electron and a neutral species.  The grains have a radius and bulk density of $a_\text{g} = 0.1 \mu$m and $\rho_\text{b} = 3$ g~cm$^{-3}$, respectively \citep{Pollack+1994}.  

\subsection{Numerical methods}
To perform our simulations, we use the 3D smoothed particle hydrodynamics (SPH) code {\sc sphNG} with the inclusion of self-gravity, radiative hydrodynamics and non-ideal MHD; this code is based upon the original version by \citet{Benz1990} and \citet{Benz+1990}, but has since been substantially modified by \citet{BateBonnellPrice1995} and many additional contributors.  

For a review of the discretised MHD equations, see \citet{Price2012}.  Briefly, we adopt the usual cubic spline kernel, set such that the smoothing length is given by $h = 1.2\left(m/\rho\right)^{1/3}$, where $m$ is the mass of an SPH particle; this yields $N_\text{neigh} \sim 58$ neighbours in three dimensions.  We calculate the gravitational forces following \citet{PriceMonaghan2007} at short range, and use a binary tree to compute the long range gravitational interactions.  For magnetic stability, the \citet{BorveOmangTrulsen2001} source-term approach is used, and artificial resistivity is included to capture the magnetic discontinuities (\citealp{PriceMonaghan2005}; \citealp{Price2012}); the artificial resistivity parameter is given by $\alpha_\text{B} = \max\left(h\left|\nabla\bm{B}\right|/\left|\bm{B}\right|,1\right)$ \citep{TriccoPrice2013}.  We employ the constrained hyperbolic divergence cleaning algorithm of \citet{TriccoPrice2012} and \citet{TriccoPriceBate2016} to control divergence errors in the magnetic field.  

To model radiation transport, we use the flux-limited diffusion method described in \citet{BateTriccoPrice2014}, where the method is described in detail in \citet{WhitehouseBateMonaghan2005} and \citet{WhitehouseBate2006}.  We use an ideal gas equation of state that assumes a 3:1 mix of ortho- and para-hydrogen \citep[see][]{Boley+2007} and treats the dissociation of molecular hydrogen and the ionisations of hydrogen and helium.  At low temperatures, the mean molecular weight is taken to be $\mu = 2.38$, and we use opacity tables from \citet{PollackMckayChristofferson1985} and \citet{Ferguson+2005}.

The code is parallelised using both OpenMP and MPI.  It does not include super-timestepping \citep{AlexiadesAmiezGremaud1996} as used in the non-ideal MHD studies by \citet{WursterPriceBate2016,WursterPriceBate2017,WursterBatePrice2018ion}, nor implicit timestepping for Ohmic resistivity as introduced in \citet{WursterBatePrice2018sd}.

\section{Initial conditions}
\label{sec:ic} 
We begin with a Bonnor-Ebert sphere \citep{Bonnor1956,Ebert1955} of radius $R=1.3\times~10^{17}$~cm, mass $M=1$~M$_\odot$, temperature $T = 8$K, and concentration parameter $\xi = 7.45$; this corresponds to a density ratio of 20:1 between the inner and outer regions of the sphere.  The ratio of thermal energy to gravitational potential energy is $\alpha = 0.50$.  The sphere is given an initial solid-body rotation $\bm{\Omega}_0 = \Omega_0\hat{\bm{z}}$, and threaded with a uniform magnetic field $\bm{B}_0$.

The sphere is placed in a low-density box of edge length $l = 4R$ at a density ratio of 1:30 with the edge of the sphere; the sphere and low-density medium are in pressure equilibrium which prevents the sphere from artificially expanding into the low-density medium.  This two-medium set-up allows us to place boundary conditions at the edge of the box rather than the edge of the sphere.  This is especially useful for the magnetic field, which is uniform initially and, therefore, we can use periodic boundary conditions at the edges of the box.  We use quasi-periodic boundary conditions at the edge of the box, in which SPH particles interact hydrodynamically `across the box', but not gravitationally.  The box size was chosen to prevent any boundary effects from influencing the evolution of the sphere.

Sink particles \citep{BateBonnellPrice1995} of radius 1 au are unconditionally inserted when the maximum density reaches $\rho_\text{crit} = 5\times 10^{-10}$ g cm$^{-3}$, and we permit only one sink particle to form per simulation.  Given the small timesteps required to evolve non-ideal MHD at high densities, the introduction of sink particles is necessary to follow the evolution of the disc long enough to determine its stability.  However, we note that sink particles stabilise small discs against instabilities \citep{MachidaInutsukaMatsumoto2014}, thus the final classification of models with small discs may be influenced by the sink particle.

Our simulations include $10^6$ SPH particles in the sphere and an additional $1.8\times 10^5$ particles in the low-density medium.  Resolving the Jeans length throughout the collapse requires at least $3\times 10^4$ particles per solar mass \citep{BateBurkert1997}, thus the Jeans mass is well resolved at all times.  The equal-mass particles in the sphere are initially placed on a regular close-packed lattice, which is then deformed to produce the Bonnor-Ebert sphere described above; the SPH particles in the warm medium have the same mass as the particles in the sphere and are also placed on a regular close-packed lattice.

We characterise the initial rotation by the orbital rotation at the outer radius of the cloud, $\Omega_\text{orbit} = 2.45\times 10^{-13}$ rad s$^{-1}$.  The five rotation speeds we test are $\Omega_0 =$ 0.05, 0.25, 0.45, 0.65 and 0.85~$\Omega_\text{orbit}$, which correspond to ratios of rotational energy to gravitational potential energy of $\beta_\text{r} = 4.4\times10^{-4}$, 0.011, 0.035, 0.074 and 0.13.  We are thus primarily exploring the higher end of the distribution of $\beta_\text{r}$-values \citep{Goodman+1993}, but the faster rotators are more likely to fragment than slower rotators. 

Our suite of models consists of purely hydrodynamical models, ideal and non-ideal MHD models.  For the magnetised models, we characterise the magnetic field in terms of the normalised mass-to-flux parameter
\begin{equation}
\label{eq:masstofluxmu}
\mu \equiv \frac{M/\Phi_\text{B}}{\left(M/\Phi_\text{B}\right)_\text{crit}},
\end{equation}
where $M/\Phi_\text{B} \equiv M/\left(\pi R^2 B\right)$ is the mass-to-flux ratio and $\left(M/\Phi_\text{B}\right)_\text{crit} = c_1/(3\pi)\sqrt{ 5/G }$ is the critical value in CGS units where magnetic fields prevent gravitational collapse altogether; here, $\Phi_\text{B}$ is the magnetic flux threading the surface of the (spherical) cloud at radius $R$ assuming a uniform magnetic field of strength $B$, and $c_1 \simeq 0.53$ is a parameter numerically determined by \citet{MouschoviasSpitzer1976}.  In this study, we test initial values of $\mu_0 =$ 3, 5, 10 and 20, which correspond to magnetic field strengths of $B_0 =$ 25.6, 15.4, 7.69 and 3.85~$\mu$G, respectively. 

In our ideal MHD models, we test two magnetic field orientations: $\bm{B}_0 = \bm{B}_\text{-z} \equiv -B_0\hat{\bm{z}}$ (i.e. the magnetic field is initially parallel to the rotation axis), and $\bm{B}_0 = \bm{B}_\text{-x} \equiv -B_0\hat{\bm{x}}$ (i.e. the magnetic field is initially perpendicular to the rotation axis).  In non-ideal MHD, the Hall effect is dependent on $\bm{\Omega} \cdot \bm{B}$ \citep{BraidingWardle2012accretion}, thus we test $\bm{B}_0 = \bm{B}_\text{+z} \equiv B_0\hat{\bm{z}}$ in addition to \bmz \ and \bmx.  We only test one orientation perpendicular to the axis of rotation (i.e. \bmx) since $\bm{\Omega} \cdot \bm{B}_{\pm\text{x}} = 0$, suggesting similar results should be obtained for \bbpmx.  However, the binary formation study of \citet{WursterPriceBate2017} found slightly different results due to the structures that formed as the systems evolved.

Due to the large suite and computational limitations, the simulations are run until the disc classification can be determined or until $t_\text{end}-t_\text{disc}\approx16$~kyr (i.e. \appx16~kyr after formation of the disc).  Models with small, dense, strongly magnetised discs typically have the shortest end-time relative to the disc formation time (aside from those models whose discs dissipate) due to the short timestep required to resolve the processes; this timestep is decreased even more in the models with the non-ideal MHD processes since they require an even shorter timestep \citepeg{Maclow+1995,ChoiKimWiita2009,WursterPriceAyliffe2014}.

\section{Results}
\label{sec:results}
In this section, we present the results of our suite of 105 models, and in Section~\ref{sec:discussion} we discuss their implications.  Details and properties of the outcome of each model (i.e., classification, disc formation time, simulation end-time, disc radius, disc mass, stellar mass, outflow and envelope properties) are listed in Tables~ \ref{table:app:results:HI} and \ref{table:app:results:N} in Appendix~\ref{app:results}.  

Our magnetised models use the naming convention of \mmodel{b}{c}{\emph{d}}{\emph{a}}, where $a=$ I (N) for ideal (non-ideal) MHD, $b$ represents 100$\times$ the initial angular rotation in terms of $\Omega_\text{orbit}$, $c$ represent the initial mass-to-flux ratio in units of the critical mass-to-flux ratio $\mu_0$, and $d$ represents the orientation of the initial magnetic field ($\pm z$ or $-x$); our hydrodynamic models are named \hmodel{b}.  An asterisk, *, in place of a variable indicates every model with the remaining defined components.

\subsection{Identifying and classifying discs}
\label{sec:results:identify:disc}

We define the total disc radius, $R_\text{T,disc}$, as the radius which includes all the gas that satisfies $\rho > \rho_\text{thresh}$, where $\rho_\text{thresh} = 10^{-13}$~\gpercc \ which is approximately the density at which the collapsing gas becomes adiabatic; this radius will include the spiral arms and the gaps between them.  The total mass of the disc, $M_\text{T,disc}$, includes all the gas with $\rho > \rho_\text{thresh}$.

To define the `bulk' disc, we divide the total disc into tori of width 0.5~au and set the height to include all the gas with $\rho > \rho_\text{thresh}$.  The bulk radius, $R_\text{B,disc}$, is then defined as the outer extent of the outermost tori where 80 per cent of the gas particles in the torus have $\rho > \rho_\text{thresh}$.  The mass of the bulk disc is the total mass of the gas with $\rho > \rho_\text{thresh}$ within the bulk radius.  For this disc to `exist', $R_\text{B,disc} \ge R_\text{crit,disc}$ and $M_\text{B,disc} \ge M_\text{crit,disc}$, where $R_\text{crit,disc} = 1$~au is the radius of the sink particle and $M_\text{crit,disc} = 6\times10^{-4}$~\Msun.  This is the disc that we will typically be referencing.   The formation time of the disc is the earliest time when the disc `exists', as per these criteria.  

We do not include any velocity or pressure criteria in this definition, thus we cannot be certain that these discs are rotationally supported.  Thus, when we refer to discs throughout this study, we are actually referring to `disc-like structures.'

We define four classifications for our models: transient, a smooth disc, spiral arms without fragmentation, and a fragmented disc.
The `transient' classification is given when a disc dissipates, i.e. when $R_\text{B,disc} < R_\text{crit,disc}$ or $M_\text{B,disc} < M_\text{crit,disc}$ occurs after a disc has formed.  Once a disc dissipates, we end the simulation, thus it is possible that a disc may re-form at a later time, but this is out of the scope of our study. 
The classifications of `smooth' and `spiral arms' are given to models that maintain a disc and do not fragment; this classification is given by visual inspection.  

\subsection{Identifying fragments}
\label{sec:results:identify:frag}

There are many ways to define a fragment, including searching for minima in the gravitational potential \citep[e.g.][]{SmithClarkBonnell2009}, finding density maxima \sm3 dex greater than their surroundings \citep[e.g.][]{MeruBate2010}, or using density gradients \citep[e.g.][]{HallForganRice2017}.  \citet{HallForganRice2017} found that the gravitational potential method was more restrictive than the density gradient method and found only clumps that survived for long periods of time, whereas the density gradient method found more clumps, including those that later merged or were accreted by the star.  We tested several methods of locating fragments and all methods found similar formation times for any given clump, within an uncertainty of $\pm\Delta t \approx 0.36$~kyr.  

In our study, we locate fragments using a method analogous to determining when sink particles are to be inserted.   We find the densest particle with $\rho > 10^{-11}$~\gpercc \ that is not associated with the central core.  We then calculate $\bm{\nabla} \cdot \bm{v}$ of this particle using the particles within 2~au of it, and we further calculate the total energy $E$ of these particles.  If $\bm{\nabla} \cdot \bm{v} < 0$ and $E  < 0$, then we define the clump as a fragment.

Although the fragmentation time is similar between our various methods, the fragmentation radius has larger variability.  The reason is that fragments are typically formed in the outer regions of the discs where the gas has not yet formed a stable orbit.  Thus the fragments and their progenitor gas radially migrate, and a slight change in our fragmentation criteria will thus have a direct impact on the fragmentation distance.  Although we present fragmentation times and distances, we do so with caution, and caution against a rigorous quantitative analysis due to these large uncertainties.

 An analysis of the evolution of the fragments is beyond the scope of this study.

\subsection{General trends}

\begin{figure*}
\begin{center}
\includegraphics[width=\textwidth]{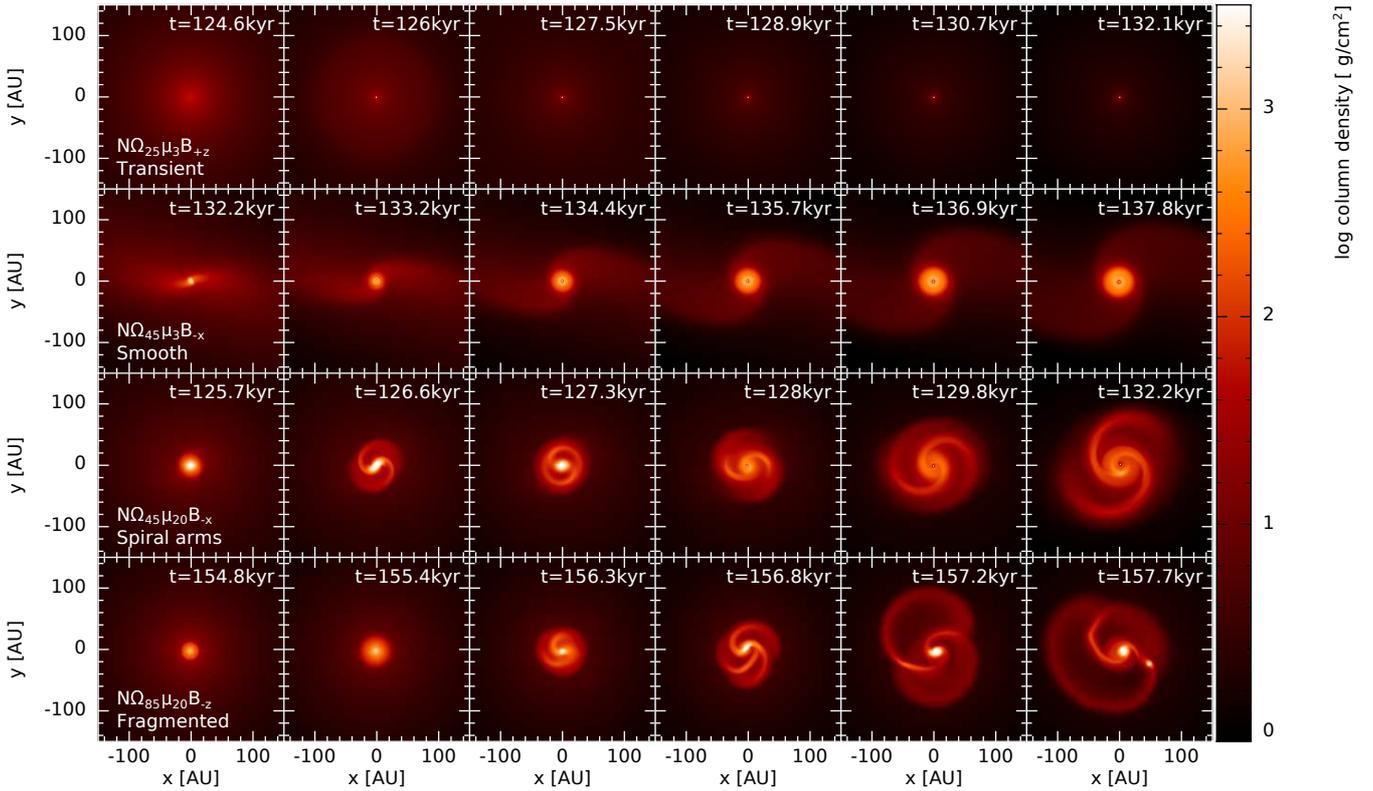}
\caption{Evolution of the gas column density of representative models that do not form a resolved disc (i.e. transient; top row), that form a smooth disc (second row), that form a disc with spiral arms that does not fragment (third row) and whose disc fragments (bottom row).  The model names are listed in the first column.  The frames are chosen to highlight evolution, and are not taken at any regular interval.  The white circle in the centre of the disc in the top three rows represents the sink particle, with the circle's radius being equal to the accretion radius of the sink.  The disc in the smooth model slowly increases in radius and decreases in surface density.   In the fragmenting model, the fragment forms at $t \approx157.4 $ kyr at a distance of $r \approx 52$ au from the core, and is visible in the final frame.}
\label{fig:results:gascolden}
\end{center}
\end{figure*}

\begin{figure*}
\begin{center}
\includegraphics[width=\textwidth]{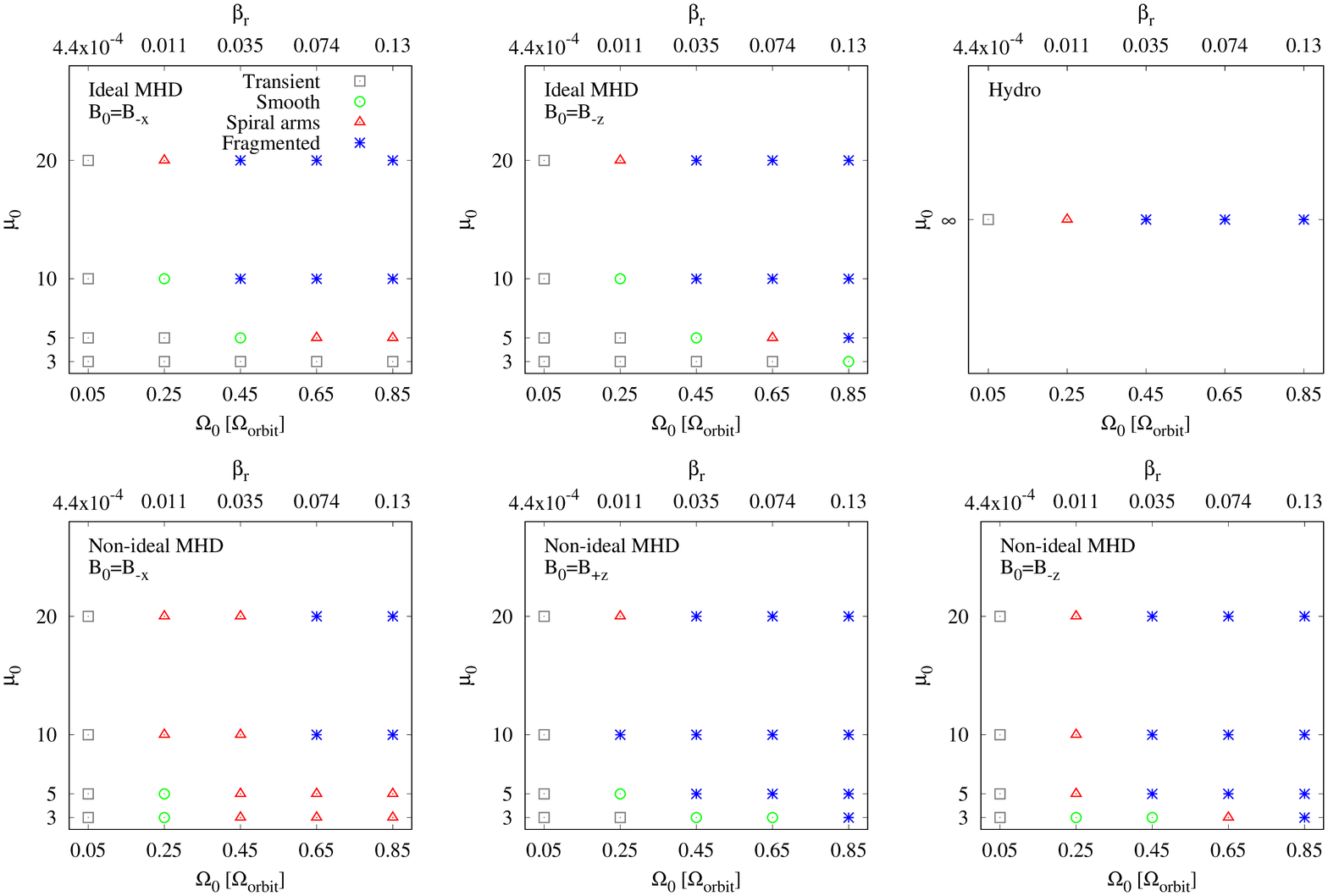}
\caption{A graphical summary of the final classification of our models.  The likelihood of fragmentation increases with both increasing initial rotation and with increasing initial mass-to-flux ratio (i.e. decreasing the magnetic field strength).  The models with \bbpmz \ are more susceptible to fragmentation than their \bbmx \ counterparts.}
\label{fig:results:summary}
\end{center}
\end{figure*}

Representative examples of the classifications are shown in Fig.~\ref{fig:results:gascolden}.   A graphical summary of the final classification of each model is given in Fig.~\ref{fig:results:summary}.
Our suite contains 105 models, of which 31 do not form discs, 12 form smooth discs, 21 form spiral arms but do not fragment, and the remaining 41 fragment.

All models with \omegaeq{0.05} are classified as transient models.  This result is independent of all magnetic properties. 

In the hydrodynamics models, discs fragment for \omegage{0.45}.  In the magnetised models, discs fragment for fast rotations (large $\Omega_0$) and weak magnetic field strengths (large $\mu_0$).  There is a general transition from fragmented, to structured, to smooth, to transients as the rotation rate is decreased and/or the magnetic field strength is increased.

For ideal MHD, the final classifications are approximately independent of the initial direction of the magnetic field (i.e. \bmz \ vs \bmx), however, the disc and fragmentation properties are dependent on the direction.   

When including the non-ideal MHD processes, larger discs tend to form than in their ideal MHD counterparts.  For the \bmx \ models, four models fail to form a disc in non-ideal MHD compared to nine in ideal MHD; however, fewer of the non-ideal MHD discs fragment (four compared to six).  For \bmz, more discs form when using non-ideal MHD, and these discs are more likely to fragment than their ideal MHD counterparts.

For \omegage{0.45}, the classifications are the same for the non-ideal MHD models with \bpmz; the classifications are also the same for the \bpmz \ models with $\mu_0=20$.  This implies that the Hall effect is not efficient enough in these regimes to affect the classification; in both regimes, the angular momentum contribution from the Hall effect is small compared to the initial angular momentum due the large initial angular momentum or the initial weak magnetic field, respectively.   For slow rotators with strong magnetic fields, discs are more likely to form in the models with \bmz, indicating that the Hall effect is efficient here.  Previous studies that investigates the Hall effect in star formation simulations \citep{Tsukamoto+2015hall,WursterPriceBate2016,Tsukamoto+2017,WursterBatePrice2018sd,WursterBatePrice2018hd,WursterBatePrice2018ff} were initialised in this parameter space.  


\subsection{Disc formation}
There are five models that never form a disc, as per the criteria given in Section~\ref{sec:results:identify:disc}.   We examine the simulations every 0.36~kyr, thus it is likely that the disc forms and dissipates within a single timestep in these models.  Thus, for these models, we substitute the disc formation time with the sink formation time.  With the exception of four models (\imodel{25}{5}{-x}, \imodel{25}{5}{-z}, \imodel{45}{3}{-z} and  \imodel{65}{3}{-z}), the discs in all the transient models dissipate within 0.72~kyr of disc formation.

\begin{figure*}
\begin{center}
\includegraphics[width=\textwidth]{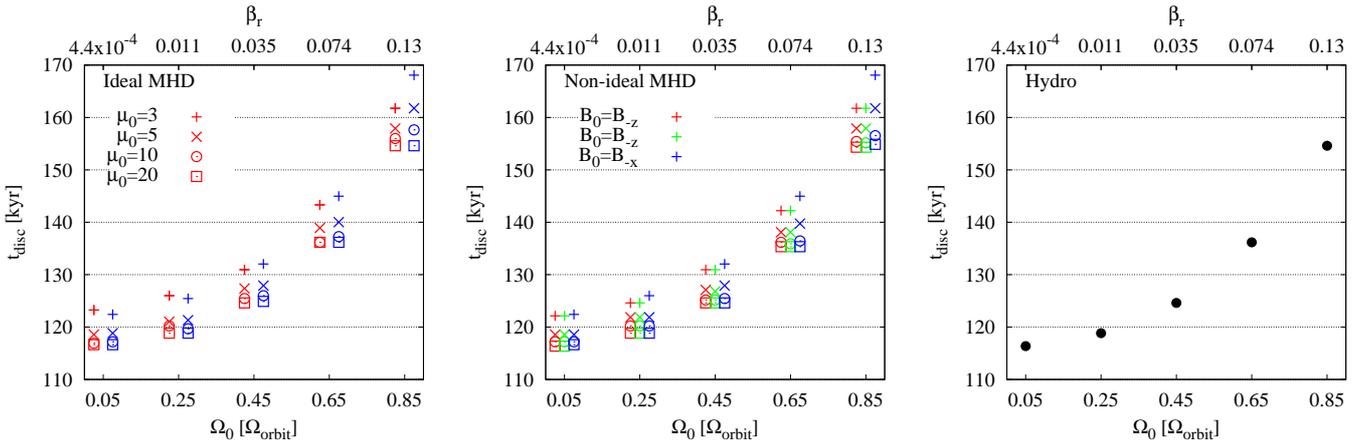}
\caption{The formation time of the discs (or the sink particle for the models that do not form a disc).  The shapes represent the initial mass-to-flux ratios and the colours represent the initial magnetic field orientation.  Values are slightly offset from the actual $\Omega_0$ for clarity.  Discs generally form later for increasing $\Omega_0$, decreasing $\mu_0$ (i.e. increasing the magnetic field strength), and switching from $\bm{B}_{\pm \text{z}} \rightarrow \bm{B}_\text{-x}$.}
\label{fig:results:form} 
\end{center}
\end{figure*}

Fig.~\ref{fig:results:form} shows the formation time of the discs.  The shapes represent the initial mass-to-flux ratios and the colours represent the initial magnetic field orientation; the initial rotational velocities are off-set from their actual value for clarity.

Discs form later for increasing $\Omega_0$ and decreasing $\mu_0$ (i.e. increasing $B_0$).  At a given $\Omega_0$ and $\mu_0$, discs tend to form at similar times for \bpmz \ and slightly later for \bmx. 

The disc formation time is approximately independent of the non-ideal MHD processes; during the initial collapse, the gas is not dense enough nor is the magnetic field strong enough for the non-ideal MHD processes to significantly affect the evolution of the system. The discs in the hydrodynamical models tend to form at similar times to the weak field magnetic models.  

These trends are expected since magnetic fields and angular momentum both support against collapse, thus increasing these values naturally leads to a later disc formation time.  When the magnetic field is initially perpendicular to the axis of rotation (i.e. \bmx), it resists the vertical collapse.  As the vertical collapse proceeds, it drags the magnetic field to the mid-plane, where the magnetic field is further amplified.  Both resisting the vertical collapse and the stronger mid-plane magnetic field strength given similar initial conditions results in the models with \bmx \ forming discs after those with \bpmz.  

In many previous studies \citepeg{PriceBate2007,HennebelleFromang2008,DuffinPudritz2009,HennebelleCiardi2009,Commercon+2010,Seifried+2011,WursterPriceBate2016}, discs failed to form in models that included ideal MHD and strong magnetic fields, consistent with the magnetic breaking catastrophe \citep{AllenLiShu2003}.  However, most of our models form discs, including 24 of 40 of our ideal MHD models.

To test for fragmentation, this study intentionally investigates a broad range of initial rotation rates, including the high-end tail as empirically determined by \citet{Goodman+1993}.  Previous studies performed fewer simulations, and were initialised with values more indicative of the mean observed rotation rates $\beta_\text{r} \lesssim 0.02$ and observed magnetic field strengths, $\mu_0 \sim 2 - 5$.  In this small parameter space, we find only one non-transient disc (\imodel{25}{5}{-x}) when using ideal MHD.  Thus, our results are consistent with the literature in the typically explored parameter space.

With ideal MHD, all of the models initially with more magnetic energy than rotational energy (i.e. $E_\text{0,mag} > E_\text{0,rot}$) either form transient or small smooth discs, whereas models with  $E_\text{0,mag} < E_\text{0,rot}$ form discs that form spiral arms or fragment.  This relation does not hold for non-ideal MHD models. 

In summary, increasing the initial rotation has the strongest effect on delaying disc formation, followed by increasing the magnetic field strength, and finally by changing the initial orientation of the magnetic field. 

\subsection{Disc fragmentation}
\label{sec:general:frag}

\begin{figure*}
\begin{center}
\includegraphics[width=\textwidth]{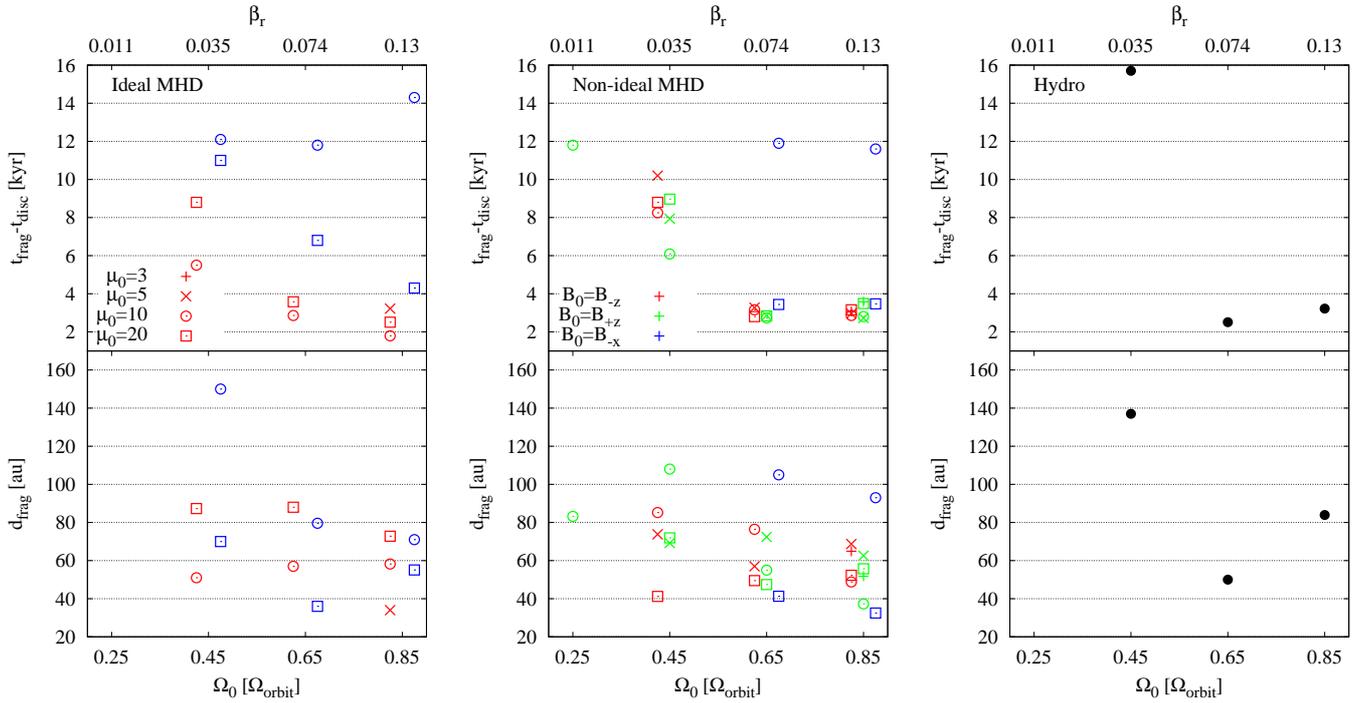} 
\caption{The fragmentation time of the discs relative to the disc formation time (top row), and the distance from the sink particle at which the fragment forms (bottom row).  No model with \omegaeq{0.05} fragments.  Recall that there are only five hydrodynamics models, three of which fragment.  There are no strong trends in either the formation time nor the formation distance, suggesting that when a disc fragments is almost independent of the initial conditions.  51 per cent of the models that fragment do so within 4~kyr after disc formation, and 78 per cent of the fragments form within 80~au of the centre of the disc.}
\label{fig:results:frag}
\end{center}
\end{figure*}

The top panels in Fig.~\ref{fig:results:frag} show the formation time of the first fragment relative to the formation time of the disc, and the bottom panels show the radius at which the fragmentation occurred. 

We find that 56 per cent of the models that fragment do so within 4~kyr after disc formation, including all but two of the non-ideal MHD models with \omegage{0.65}.  However, there is no trend of fragmentation time with respect to any of our parameters.  Of the models with \omegaeq{0.45} that fragment, they do so $ > 5$~kyr after disc formation.  These discs are initially compact, and remain small in the presence of strong magnetic fields (either low-$\mu_0$ or \bmx).  Thus, compared to models with weaker magnetic fields which quickly form large discs, these models require additional time to grow in radius and for the magnetic Toomre-Q parameter, $Q_\text{m}$, to decrease into the instability regime where it may then fragment (see Section~\ref{sec:TQ} below).

As with the fragmentation time, there is no strong correlation between fragmentation distance and our parameters, although models with lower $\Omega_0$ tend to fragment at further distances from the centre.  Approximately 78 per cent of the models that fragment do so at initial distances of $r \lesssim 80$~au from the centre of the disc.   Prior to fragmentation, the disc becomes unstable and forms spiral arms, and it is in the arms where the fragmentation occurs; in most cases, the fragment forms in the middle of the arm rather than near the bulk disc or near the tip.

The fragmentation distance must be taken with caution since the fragments typically migrate during and after formation, which leads to a large uncertainty about their specific formation distance \citepeg{KleyNelson2012,Baruteau+2014,Meru+2019}.  Like the fragmentation times, the fragmentation distance is independent of the initial parameters in our suite.

\subsection{Disc Properties}
\label{sec:disc:disc}
\begin{figure*}
\begin{center}
\includegraphics[width=\textwidth]{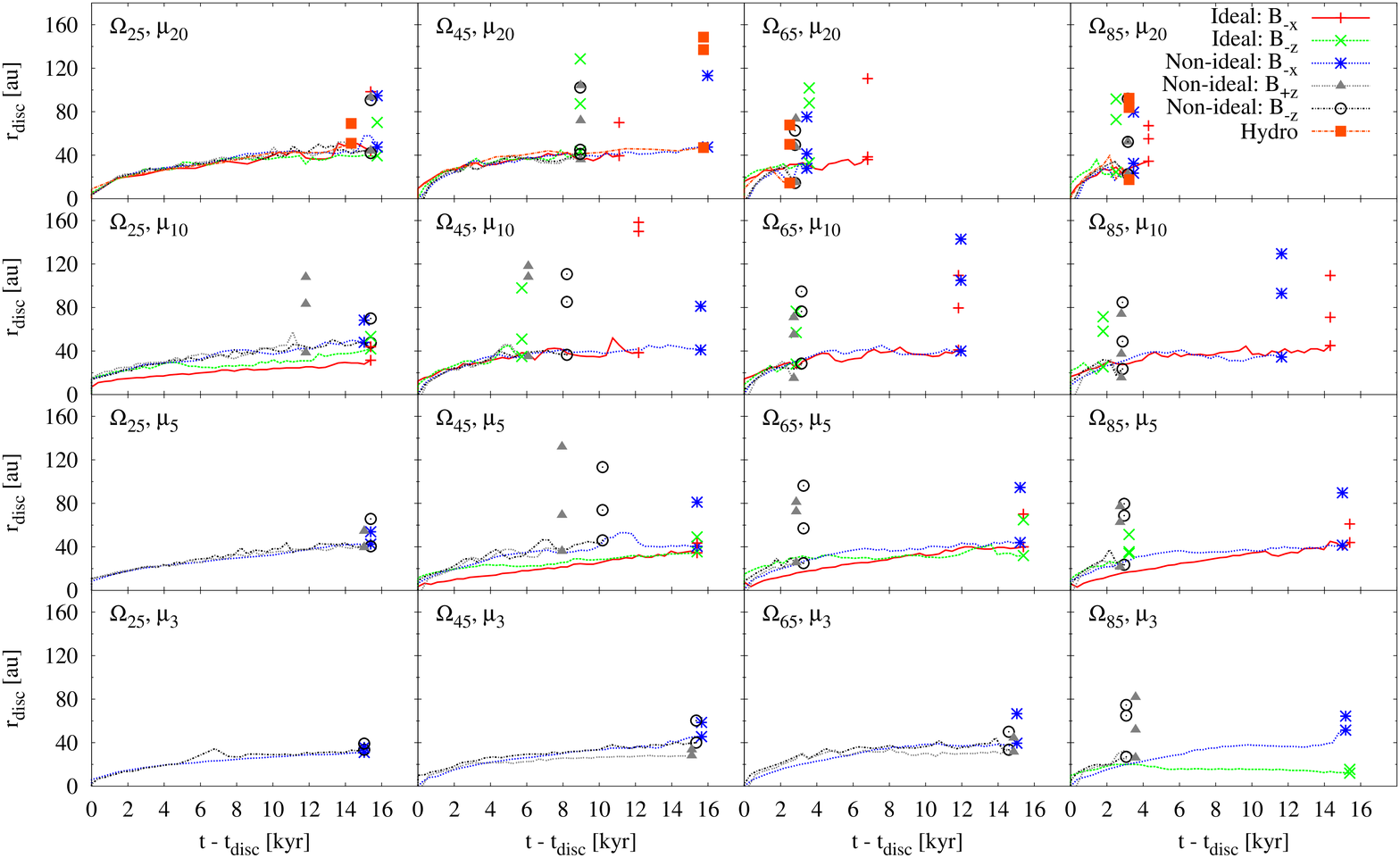} 
\caption{The evolution of the bulk radius of each non-transient model in our suite (lines).   At the end of each line are two or three points at increasing radii but the same time, which represent the final bulk disc radius, the fragmentation radius (if it exists) and the final total radius, from smallest to largest, respectively.  The key shows both the line-style and symbol corresponding to each model, although they are used individually.  As an example, in the $\Omega_{45}$, $\mu_5$ panel, models \nmodel{45}{5}{$\pm$z} fragment, thus the three triangles represent the bulk disc radius, fragmentation radius and total radius of \nmodel{45}{5}{+z} at the time it fragmented, while the three circles show the same for \nmodel{45}{5}{-z}; the remaining three models  do not fragment, thus at $t - t_\text{disc} \approx 16$~kyr, the symbols represent the bulk and total radius of these three models.  The two points for \imodel{45}{20}{-x} are bulk radius and fragmentation radius; the final bulk radius is at 200~au. Disc radii increase with time; the dependence on their growth rate on the initial conditions becomes stronger for models with stronger initial magnetic fields.  Models with faster initial rotations typically have larger total radii, and the majority of the models that fragment form their fragment closer to the tip of the spiral arm than its base.}
\label{fig:disc:radius}
\end{center}
\end{figure*}

\begin{figure*}
\begin{center}
\includegraphics[width=\textwidth]{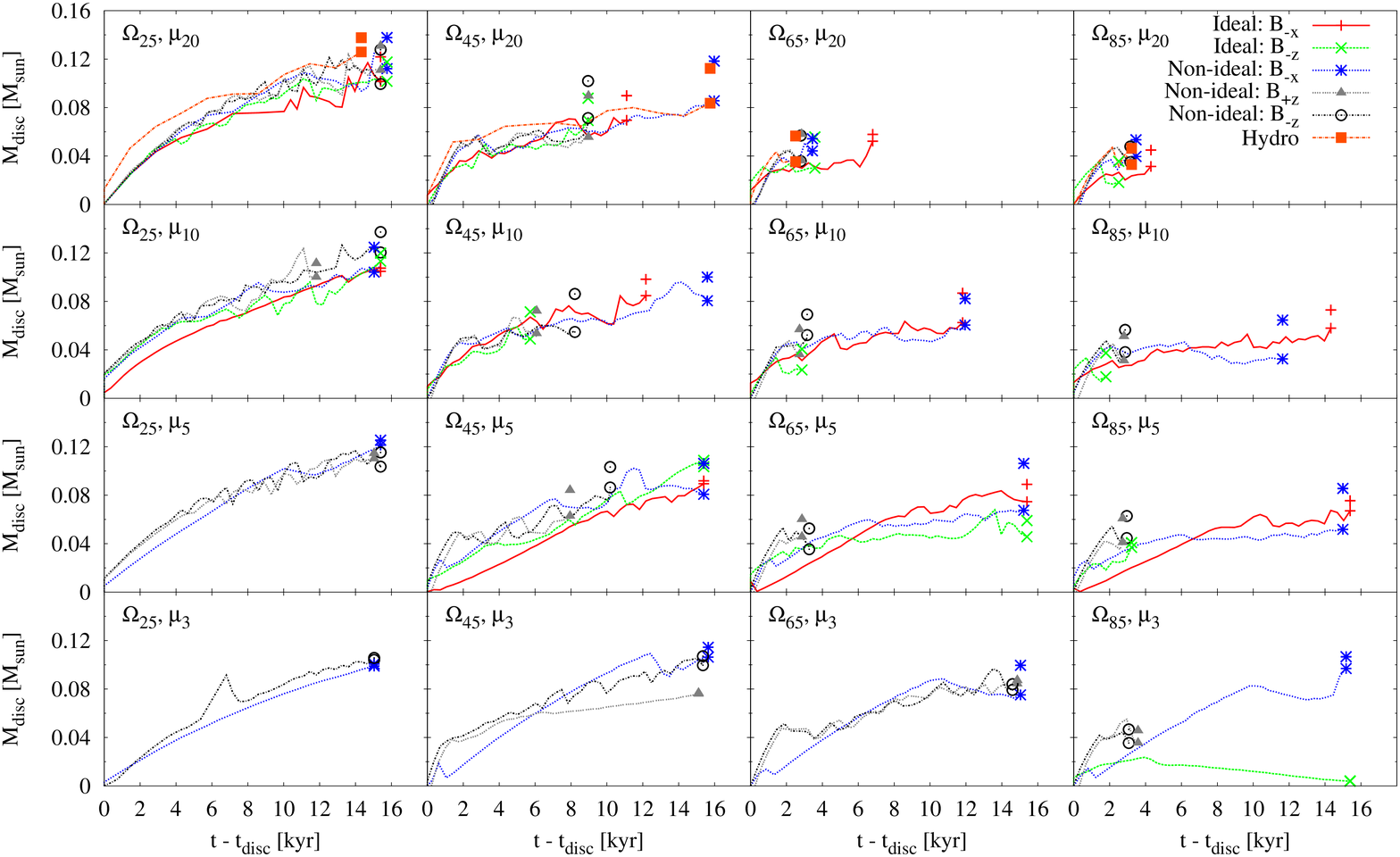} 
\caption{The evolution of the bulk discs mass of each non-transient model in our suite (lines).  At the end of each line are two points at increasing masses but at the same time, which represent the the final bulk disc mass (lower) and the final total mass (upper).  The disc masses increase with time, with a slight trend of less massive discs for models with higher initial rotation rates.}
\label{fig:disc:mass}
\end{center}
\end{figure*}

The discs are continually evolving, as shown by the time sequence in Fig.~\ref{fig:results:gascolden}.  The evolution of the bulk disc radii and bulk disc mass are shown by the lines in Figs.~\ref{fig:disc:radius} and \ref{fig:disc:mass}, respectively, with the final bulk and total values shown by the points; the final radii and masses are also given in Tables~\ref{table:app:results:HI} and \ref{table:app:results:N}.
 
In general, the bulk disc radii continue to grow with time as gas accretes on to them, with final radii of 30-50~au at $t - t_\text{disc} \approx 16$~kyr.  The models that fragment necessarily have smaller final radii since they have had less time to grow.  Models with an initially weak magnetic field typically undergo an initial rapid accretion phase followed by slower growth, and the growth rate for the \mueq{20} models is only trivially dependent on the orientation of the magnetic field or the initial rotation.  The dependence on initial conditions becomes more pronounced for stronger magnetic fields, however there are fewer non-transient models to analyse.  Where there are slight differences amongst the growth rates, the non-ideal MHD models form larger bulk discs than their ideal MHD counterparts.

Fig.~\ref{fig:disc:radius} also plots the final total radius (defined at the outermost radius where a gas particles has \rhogt{-13}, thus typically represents the radius of the tip of the spiral arm).  The total radii spans a large range, and is typically larger for models with weak magnetic fields or high rotations.  The smooth disc models typically have total radii of $r_\text{T,disc} / r_\text{B,disc} \lesssim 2$, while the models that form spiral arms and/or fragment can have $2 \lesssim r_\text{T,disc} / r_\text{B,disc} \lesssim 6$.  The large ratio and large total radius indicate the presence of substantial spiral arms (independent of whether or not they fragment).

Fig.~\ref{fig:disc:radius} also gives an indication about the location of the fragment.  Fragmentation typically occurs at $1.5 \lesssim r_\text{frag} / r_\text{B,disc} \lesssim 4$, and the majority (25 of 41) of the models fragment closer to the tip of the spiral arms rather than the edge of the bulk disc; however, three models (\nmodel{45}{20}{-z}, \imodel{65}{20}{-x} and \imodel{85}{5}{-z}) fragment on the edge of the bulk disc.  Thus, for fragmentation to occur, the prior formation of spiral arms is clearly beneficial.

The disc masses increase over their lifetime, indicating that they are continually accreting gas (Fig.~\ref{fig:disc:mass}).  These values are dependent on the accretion onto the disc, accretion onto the star from the disc, and the instantaneous calculation of the radius.  Thus, these values have noticeable variability, and it is challenging to extract meaningful trends, although there is a slight trend of lower discs masses for models with faster initial rotations.  In all cases, the majority of the gas is in the bulk disc (comparing the symbols at the end of each curve), indicating that the there is very little dense gas in the spiral arms.  The models that fragment tend to have higher fractions of mass in the spiral arms (30-40 per cent), suggesting that the fragments that form are (at least initially) low mass.

Throughout the lifetime of the disc-like structures, they are rotating at sub-Kelperian speeds.  Towards the end of the simulations, the rotational speed has increased to being only a few times slower than the Keplerian speed.

\section{Discussion}
\label{sec:discussion}

This section further discusses fragmentation, however, we also discuss some other characteristics and trends that arise as a direct result of varying our parameter space.

\subsection{Hydrodynamic versus magnetised models}
\label{sec:disc:hi}

At any given \omegaz, the purely hydrodynamic models should collapse faster than their strongly magnetised counterparts since there is no support from magnetic fields.  Indeed, the discs form at similar times to the weakly magnetised models, where the magnetic field does not play an important role in the evolution of the system.  Without magnetic fields to transport angular momentum away from the collapsing central region, rotationally supported discs form, with the size increasing from models with $\Omega_0 = 0.25$ to $0.45\Omega_\text{orbit}$; these discs are larger than their magnetised counterparts at any given \omegaz.  Models with $\Omega_0 = 0.65$ and $0.85\Omega_\text{orbit}$ quickly become unstable and fragment, thus a direct comparison of disc size is not feasible.  

Since the hydrodynamic models form discs that are larger than their strongly magnetised counterparts (assuming the discs grow rather than immediately fragment), it is reasonable to expect that if a hydrodynamic model does not fragment, then neither will its magnetised counterparts.  This is true, with the exception of \nmodel{25}{10}{+z}.  The disc in \hmodel{25} is larger and more extended, with wide spiral arms close to the bulk disc.  The bulk disc in \nmodel{25}{10}{+z} is condensed, with a narrow, extended arm in near isolation.  These narrow spiral arms are typically more susceptible to instability and fragmentation than wider arms, frequently forming over-densities.  In some cases, the over-densities dissipate back into the arm, but the more frequent result is fragmentation, as in the case of \nmodel{25}{10}{+z}.

\subsection{Ideal versus non-ideal MHD}
\label{sec:disc:ini}

\subsubsection{Parallel versus perpendicular magnetic field}

Generally, the discs in the \bmx \ models tend to be slightly larger than their \bpmz \ counterparts.   When they do fragment, it tends to be later, suggesting that this orientation of magnetic field stabilises against fragmentation, but typically delays rather than prevents it.   The mid-plane magnetic field strengths tend to be stronger in the \bmx \ models than their \bpmz \ counterparts \citep[in agreement with][]{WursterPriceBate2017}.  

\subsubsection{Parallel magnetic field}
As discussed above, discs are less likely to form in strong magnetic fields in the ideal MHD approximation \citep[e.g.][]{AllenLiShu2003}, and indeed, only our fastest rotating ideal MHD model with \mueq{3} forms a disc.  By including non-ideal MHD with the \bmz \ orientation, four of the \mueq{3} models form discs, and \nmodel{85}{3}{-z} even fragments.  By reversing the direction of the magnetic field, the Hall effect transports the angular momentum in the gas around the protostar outwards, hindering disc formation.  Thus, \nmodel{5}{3}{+z} and \nmodel{25}{3}{+z} fail to form discs.  However, the magnetic dissipation from Ohmic resistivity and ambipolar diffusion permit discs to form in \nmodel{45}{3}{+z} and \nmodel{65}{3}{+z} unlike in their ideal MHD counterparts.

At \mueq{5} and \omegage{0.45}, the discs fragment for the non-ideal MHD models but not the ideal MHD models, as a result of physical resistivity allowing larger discs to form.  The Hall effect contributes oppositely to the angular momentum in the discs for the models with \bpz \ and \bmz, however, the azimuthal ion velocity is similar to the bulk azimuthal velocity, showing that Hall effect cannot overcome the fast initial rotation to make significant changes to the evolution; the fragmentation is only slightly delayed in \nmodel{45}{5}{-z} compared to \nmodel{45}{5}{+z}.  The Hall effect is strong enough at \omegaeq{0.25} such that the disc in \nmodel{25}{5}{-z} is larger than in \nmodel{25}{5}{+z}, but neither fragment.  

In the ideal MHD models, although the magnetic field is initially \bmz, a strong toroidal component develops in the discs.  This occurs for all the ideal MHD models that form discs (except for \imodel{5}{20}{-z}).  The non-ideal MHD models also develop a toroidal component, but it is generally weaker than in their ideal counterparts.  

\subsubsection{Perpendicular magnetic field}

The discs are typically larger in the non-ideal MHD models than their ideal counterparts due to less magnetic braking.  Unlike the ideal MHD models, discs form for \omegage{0.25} and \mueq{3}.

Models \nmodel{10}{45}{-x} and \nmodel{20}{45}{-x} do not fragment, unlike their ideal MHD counterparts.  All four form narrow and dense arms, and the arms in the non-ideal MHD models are permeated with weaker magnetic fields than in their ideal counterparts, thus spread out and do not fragment.  The spiral arms in the ideal MHD models become more well defined as they evolve until they fragment $>~10$~kyr after the discs has formed.  

\subsection{Toomre-Q parameter}
\label{sec:TQ}

The Toomre-Q parameter \citep{Safronov1960,Toomre1964} is given by
\begin{equation}
\label{eq:ToomreQ}
Q = \frac{\kappa c_\text{s}}{\pi \Sigma G}
\end{equation}
where $c_\text{s}$ is the local sound speed, $\kappa$ is the epicyclic frequency and $\Sigma$ is the surface mass density.  In the presence of magnetic fields, the magnetic Toomre-Q parameter is given by 
\begin{equation}
\label{eq:ToomreQm}
Q_\text{m} = \frac{\kappa\sqrt{c_\text{s}^2+v_\text{A}^2}}{\pi \Sigma G},
\end{equation}
where $v_\text{A}$ is the Alfv{\'e}n velocity.

The epicyclic frequency is given by
\begin{equation}
\label{eq:kappa}
\kappa^2 = \frac{2\Omega}{r}\frac{\text{d}}{\text{d}r}\left(r^2\Omega\right),
\end{equation}
where $\Omega$ is the angular frequency of the disc.  For Keplerian discs, $\kappa \approx \Omega$, and this is the version that is commonly presented in the literature.  However, during disc formation, the young discs rotate with sub-Keplerian speeds, thus we calculate the Toomre-Q parameter using $\kappa$ rather than $\Omega$.  The fluid becomes unstable for $\kappa^2 < 0$. 

Discs are susceptible to fragmentation when $Q \lesssim Q_\text{crit}$.  For an infinitesimally thin hydrodynamics disc, $Q_\text{crit} \sim 1$, while for a 3D hydrodynamics disc $Q_\text{crit} \sim 1.5-1.7$ (e.g. \citealt{Durisen+2007}, \citealt{Helled+2014} and references therein).   Using 2D shearing box simulations, \citet{KimOstriker2001} determined that for magnetised self-gravitating discs, $Q_\text{m,crit} \sim 1.2-1.4$, where this range is given for the growth of non-axisymmetric perturbations (see also \citealt{KimOstrikerStone2003}).

\begin{figure}
\begin{center}
\includegraphics[width=\columnwidth]{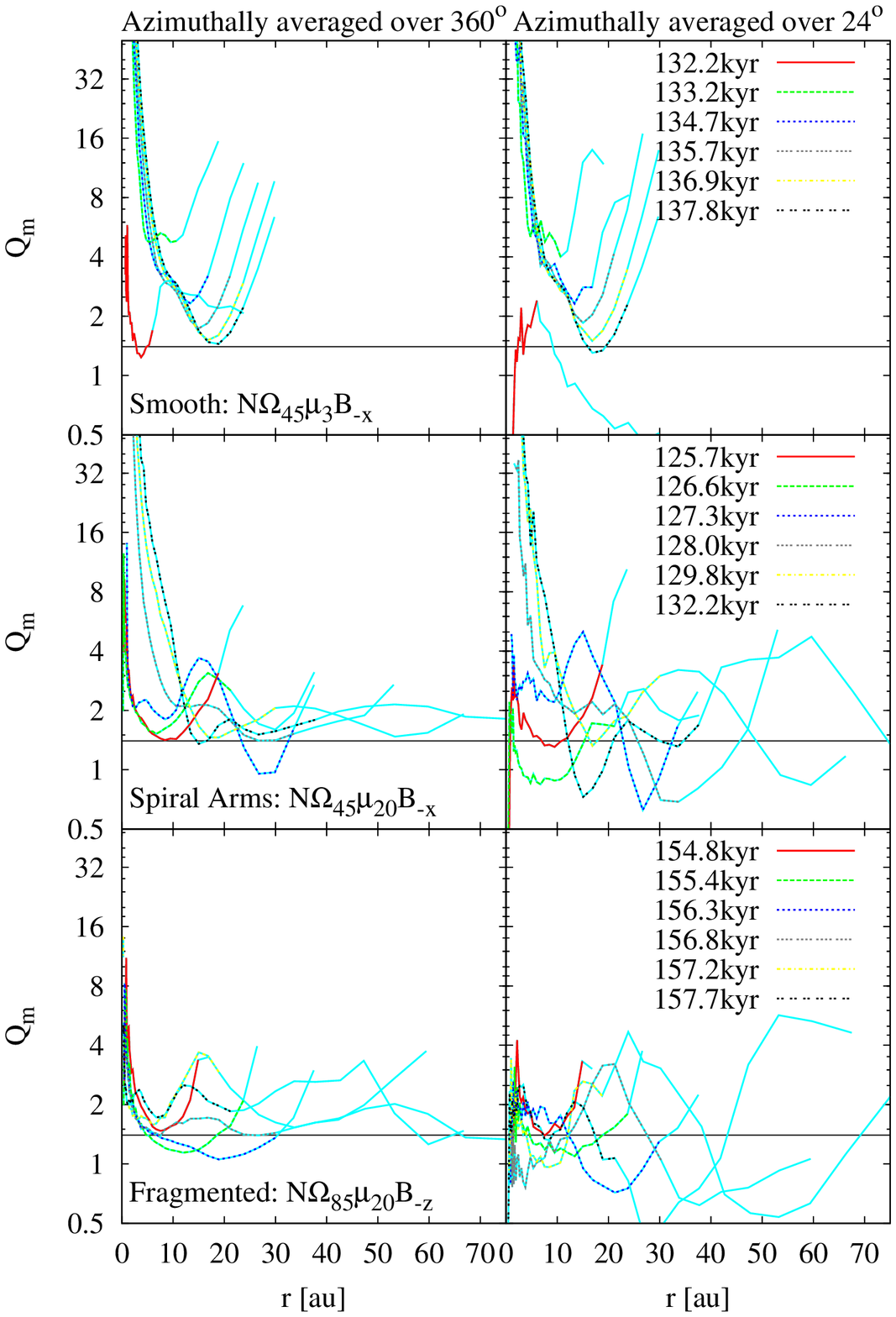}
\caption{The magnetic Toomre-Q parameter along with the 2D stability limit of $Q_\text{m,crit} \approx 1.4$ for the representative cases from Fig.~\ref{fig:results:gascolden} that form a disc.  The left-hand panel shows the azimuthally averaged $Q_\text{m}$, and the right-hand panel shows $Q_\text{m}$ of the 24$^\circ$ wedge that contains the minimum Toomre-Q value, $Q_\text{m,min}$; the times are as listed in the figure.  The coloured segment of the line represents the bulk discs, and the cyan line segments extend to the total disc radius.
Both \nmodel{45}{20}{-x} and \nmodel{85}{20}{-z} have $Q_\text{m,min} < 1.4$, yet only the latter model fragments.  This suggests that $Q_\text{m}$ alone is not sufficient to determine if a disc will fragment.}
\label{fig:results:Qm}
\end{center}
\end{figure}

The left-hand column of Fig.~\ref{fig:results:Qm} shows the azimuthally averaged $Q_\text{m}$ at six times for the three representative cases shown in Fig.~\ref{fig:results:gascolden} that form discs.  Given the asymmetric nature of many of our discs, we also calculate $Q_\text{m}$ in wedges of $24^\circ$ to search for local minima that may not be detectable in the azimuthally averaged values; see the right-hand column of Fig.~\ref{fig:results:Qm} for the $Q_\text{m}$ wedge that contains the minimum value.

In our suite of models, $Q_\text{m}$ yields limited insight into disc fragmentation.  As our smooth example disc evolves  (top panel of Fig.~\ref{fig:results:Qm}), the minimum $Q_\text{m}$, given by $Q_\text{m,min}$, slowly decreases.   When considering the wedge, the values are slightly lower suggesting a slight asymmetry in the disc, but there is still no indication that this disc will fragment (visually confirmed in the second row of Fig.~\ref{fig:results:gascolden}).

Both \nmodel{45}{20}{-x} (middle panels) and  \nmodel{85}{20}{-z} (bottom panels) have $Q_\text{m,min} < 1.4$ in the global (azimuthally averaged) and wedge profiles.  The global value in the former only briefly drops below the 2D stability limit, suggesting that the over-density quickly diffuses rather than collapses and fragments.  The global value in the latter is only marginally unstable at the radius where the disc ultimately fragments.  When considering the wedges, both models have several regions that are Toomre-unstable, which suggest that both models have regions that are susceptible to fragmentation.  However, only the latter model fragments.  Thus, we cannot clearly determine the outcome of a model based upon $Q_\text{m}$ alone.

Contrary to the top panel of Fig.~\ref{fig:results:Qm}, many of the stable disc models yield regions of $Q_\text{m} < 1.4$ but do not fragment.  Moreover, most of the discs in our suite have regions that are Toomre-unstable, even those that form smooth discs or discs with spiral arms.  Thus, we conclude that all models that fragment do so in regions with $Q_\text{m} < 1.4$, but not all regions with $Q_\text{m} < 1.4$ necessarily fragment.  

\begin{figure}
\begin{center}
\includegraphics[width=\columnwidth]{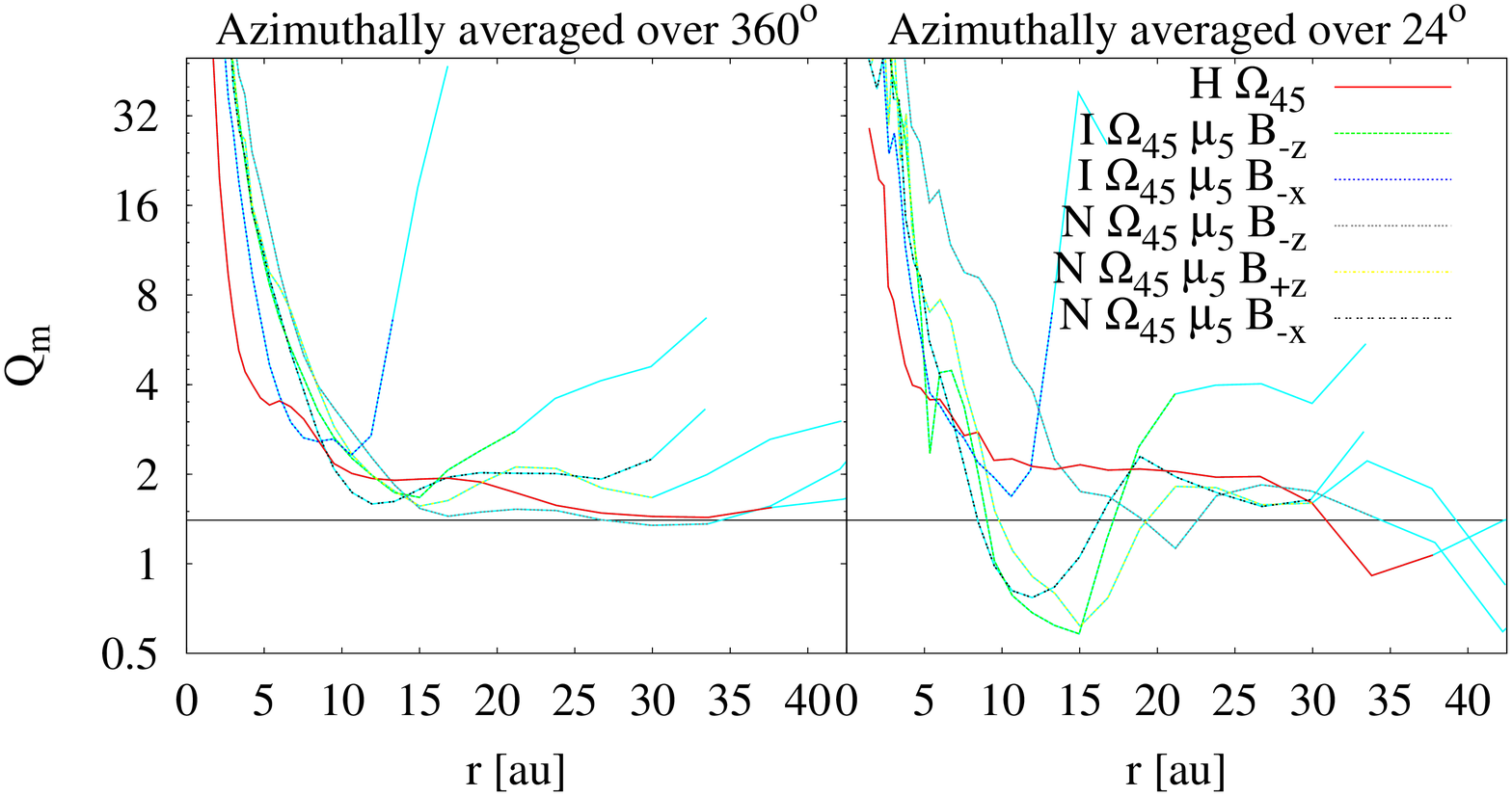}
\caption{The magnetic Toomre-Q parameter along with the 2D stability limit of $Q_\text{m,crit} \approx 1.4$ at 4~kyr after disc formation for each model with \mueq{5} and \omegaeq{0.45}.  The left-hand panel shows the azimuthally averaged $Q_\text{m}$, and the right-hand panel shows $Q_\text{m}$ of the 24$^\circ$ wedge that contains $Q_\text{m,min}$.  At this time, the azimuthally average values suggest stability, while over-dense regions are identified by the wedge-values.  These over-densities will ultimately disperse into the discs; models \hmodel{45} and \nmodel{45}{5}{$\pm$ z} ultimately fragment at $\gtrsim 8$~kyr after disc formation, while the remaining models do not. }
\label{fig:results:Qm4}
\end{center}
\end{figure}

\begin{figure}
\begin{center}
\includegraphics[width=\columnwidth]{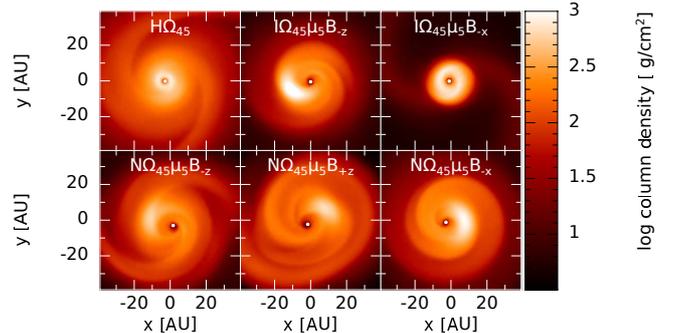}
\caption{Gas column density of the models with \omegaeq{0.45} and \mueq{5} at 4~kyr years after disc formation as analysed in Fig.~\ref{fig:results:Qm4}.  The disc size is directly dependent on the orientation of the magnetic field and the inclusion of the non-ideal MHD processes.   Asymmetries and transient  local over-densities appear in most models. }
\label{fig:results:mu5O45}
\end{center}
\end{figure}

To compare the effect of the magnetic field orientation, Fig.~\ref{fig:results:Qm4} shows $Q_\text{m}$ for the magnetised models with  \mueq{5} and \omegaeq{0.45} 4~kyr after disc formation; the gas column density of these models at this time is show in Fig.~\ref{fig:results:mu5O45}.

At this time, these models have formed rotationally supported discs.  The azimuthally averaged $Q_\text{m}$ suggests that these discs are stable, however, the wedge values suggest there are unstable regions for each model except \imodel{45}{5}{-x}.  Each of these locations of  $Q_\text{m,min} < 1.4$ corresponds to an over-density clearly visible in Fig.~\ref{fig:results:mu5O45}, with the exception of \hmodel{45}.  These over-densities do not correspond to an increase/decrease in any other property, including gas temperature, velocity, magnetic field strength or velocity, and quickly disperse.  Models  \nmodel{45}{5}{+z}, \nmodel{45}{5}{-z}  and \hmodel{45} ultimately fragment at $t \approx 8$, 10 and 16~kyr after disc formation, respectively.

Thus, the orientation of the magnetic field clearly influences the formation and evolution of the disc, even early on.  Most of the example models form clear over-densities, but these do not fragment, again suggesting that $Q_\text{m}$ alone does not determine if a disc will fragment. 

\subsection{Disc-to-star mass ratios}
\label{sec:massratio}
Massive discs are more likely to fragment than less massive discs, and fragmentation is expected for 
\begin{equation}
\label{eq:massratio}
\frac{H}{R} \lesssim \frac{M_\text{disc}}{M_\text{star}},
\end{equation}
where $H = c_\text{s}/\Omega$ is the isothermal scale height  \citep{Gammie2001}.  

In isothermal, Keplerian discs, it is expected that $H/R \sim 0.1$.  Our discs are not isothermal, with the sound speed varying by a factor of \sm10 throughout the disc.  
Taking the average sound speed and average angular frequency over the disc, we find that $0.1 \lesssim H/R  \lesssim 0.15$ for $t > t_\text{disc} + 4$~kyr for the bulk discs, and $0.05  \lesssim H/R  \lesssim 0.1$ for the total discs.  Thus, on average, $H/R \sim 0.1$ is an appropriate approximation in our models, which is in general agreement with the literature.

For this calculation, the stellar mass, $M_\text{star}$,  is given by the mass of the sink particle of radius 1~au, and the disc mass is that of the bulk disc.  After the initial accretion phase, the mass ratio is typically ${M_\text{disc}}/{M_\text{star}} \sim 1$ (this is not true for the models with \omegaeq{0.05} and the ideal MHD models with \mueq{3}).  This is reasonable since the protostellar discs have just formed, and the central regions have just collapsed to form the protostar.  For the duration of the simulation, this ratio is approximately constant, but we cannot comment upon the long term evolution of this ratio.  

In general, ${H}/{R} \sim 0.1 {M_\text{disc}}/{M_\text{star}}$ for the models that form spiral arms or fragment, thus in these models, the condition in Eqn.~\ref{eq:massratio} is satisfied.  However, these models and many of the smooth models maintain ${H}/{R} < {M_\text{disc}}/{M_\text{star}}$ throughout the simulation, thus this relationship is a poor discriminate to determine fragmentation.

\subsection{Outflows}
\label{sec:disc:out}
Although not the main focus of this study, the large parameter space allows us to briefly investigate outflows.  Since our models use 1~au sink particles, these outflows are first core outflows \citep[for a more detailed discussion on first core outflows in non-ideal MHD simulations, see][]{WursterBatePrice2018sd,WursterBatePrice2018hd}.  

Slow ($v_\text{r} < 1$~\kms), broad outflows are launched in the ideal MHD models with \mule{5}, \omegage{0.25} and \bmz.  Slow outflows are also launched in \nmodel{25}{3}{-z} and  \nmodel{85}{3}{$\pm$ z}, while fast ($v_\text{r} \sim 1-5$~\kms) outflows are launched in  \nmodel{5}{3}{+z},  \nmodel{25}{3}{+z},  \nmodel{45}{3}{$\pm$ z},  \nmodel{65}{3}{$\pm$ z}.  Four of these ideal MHD and two of the non-ideal MHD models have transient classifications.  No outflows are launched in our models with \bmx.

From the ideal MHD models, this suggests that magnetic fields with a strong poloidal component are required in addition to a reasonable amount of initial angular momentum.  Ohmic resistivity and ambipolar diffusion weaken the magnetic field enough such that outflows are not launched in the \mueq{5} models that include the non-ideal MHD processes, and that fast outflows are launched in six of the nine non-ideal MHD models that launch outflows.  Thus, we find that outflows are dependent primarily on direction and strength of the magnetic field, where the strength is necessarily weakened by the inclusion of the non-ideal MHD processes.  We generally find outflow speeds decreasing with increasing $\Omega_0$, since larger discs are permeated by a similar magnetic flux as smaller discs which results in less magnetic pinching and weaker field strengths in the larger discs.

This result suggests a resolution of conflicting results in the literature.  Both \citet{WursterPriceBate2017} and \citet{KuruwitaFederrathIreland2017} modelled the formation of binary stars, however, only the models in \citet{KuruwitaFederrathIreland2017} launched outflows.  The system in \citet{KuruwitaFederrathIreland2017} yielded smaller binary separations and smaller discs than \citet{WursterPriceBate2017}, and outflows that carried angular momentum away from the protostars were launched.  The calculations of \citet{WursterPriceBate2017} produced large discs and the wide binary separations.  To verify that these differences were a result of the initial conditions and not a difference in the algorithm (i.e. SPH vs adaptive mesh refinement), we previously ran low-resolution proof-of-concept models using the algorithms from \citet{WursterPriceBate2017} and the initial conditions from \citet{KuruwitaFederrathIreland2017} and found that, indeed, outflows were formed.  Thus, large discs in ideal MHD simulations do not appear to launch early outflows.

\subsection{Counter-rotating envelopes}
\label{sec:disc:counter}
It has been previously shown that models that include the Hall effect and initial magnetic field orientations of \bmz \ produce counter-rotating envelopes \citep[e.g.][]{KrasnopolskyLiShang2011,LiKrasnopolskyShang2011,Tsukamoto+2015hall,WursterPriceBate2016,Tsukamoto+2017,WursterBatePrice2018ion,WursterBatePrice2018hd}.  This is to conserve angular momentum as the Hall effect spins up the gas around the protostar.  Since the Hall effect hinders disc formation in models with \bpz, these studies found no counter-rotation in models with \bpz.   

\begin{figure*}
\begin{center}
\includegraphics[width=\textwidth,trim={0 1.2cm  0 0}]{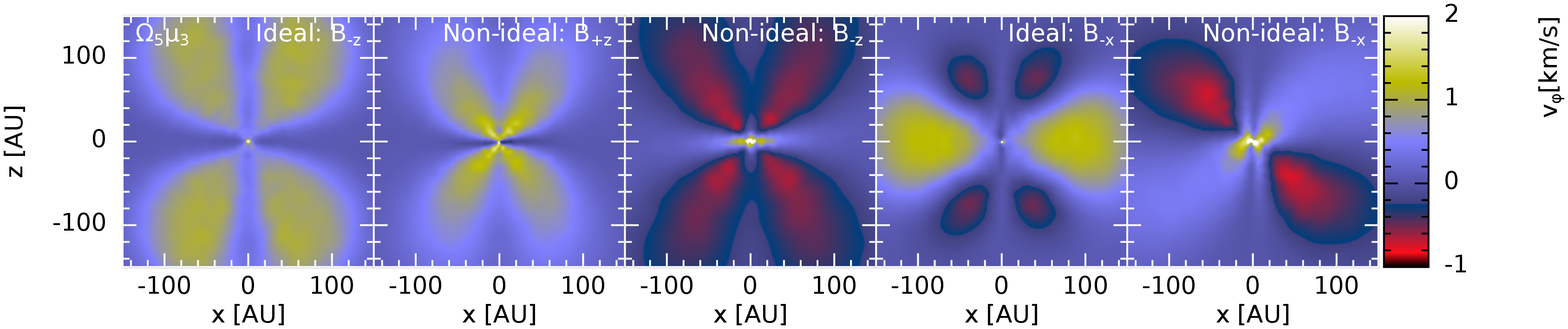}
\includegraphics[width=\textwidth]{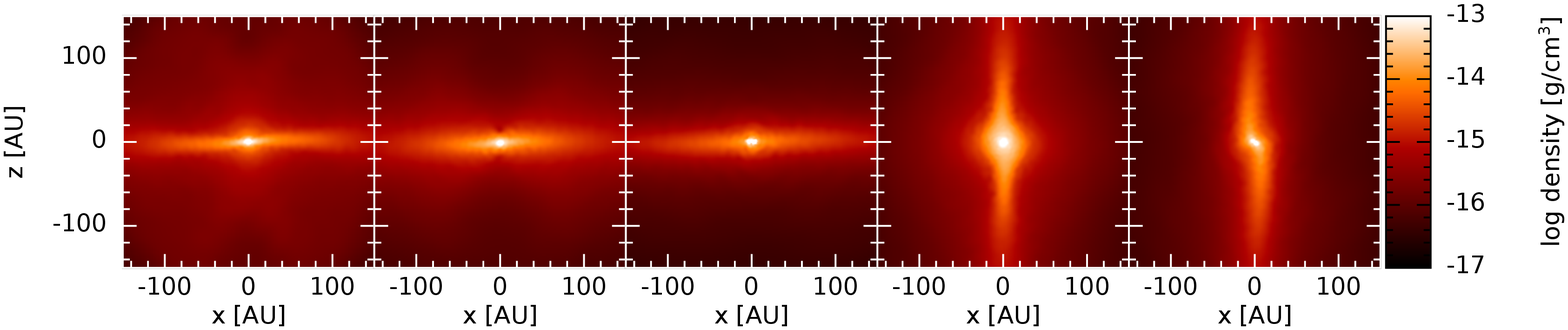}
\caption{Azimuthal velocity (top row) and gas density (bottom row) in a plane through the centre of the protostar parallel to the rotation axis near the end of the simulation for the models with \omegaeq{0.05} and \mueq{3}.  The initial rotation is $v_\phi > 0$.  Counter-rotating envelopes form in the slowly-rotating non-ideal MHD model with \bmz, but not the model with \bpz.  Model \imodel{5}{3}{-x} forms a counter-rotating envelope, which is the only ideal MHD model in our entire suite to do so.}
\label{fig:vy}
\end{center}
\end{figure*}

In our entire suite, no counter-rotating envelopes form for \omegage{0.25}, since the Hall effect is not strong enough to overcome the initial rotation of the envelope.  As expected from the previous studies, models \imodel{5}{*}{-z} and \nmodel{5}{*}{+z} do not form counter-rotating envelopes; \nmodel{5}{3}{-z} forms a strong counter rotating envelope, and \nmodel{5}{5}{-z} forms a weak one.  See the first three panels of the top row in Fig.~\ref{fig:vy}, which shows the azimuthal velocity $v_{\phi}$ and gas density of the \omegaeq{0.05} models with \mueq{5} near the end of their respective simulations; note that $v_{\phi,0} > 0$.

In the above cases, and typically discussed throughout the literature, the cause of the counter-rotating envelope is the Hall effect when the magnetic field and rotation vectors are anti-aligned.  However, counter-rotating regions may also form if the magnetic field is initially perpendicular to the rotation axis (i.e. \bmx).

A gas over-density forms along the rotation axis in \imodel{5}{3}{-x} (fourth column in Fig.~\ref{fig:vy}).  The low-density gas in the mid-plane rotates rapidly, while above and below form a slow counter-rotating envelope.  In all other ideal MHD models the initial rotation is strong enough to prevent a counter-rotating envelop from forming.  

In \nmodel{5}{3}{-x}, the Hall effect causes the gas to misalign from the rotation axis such that the normal to the dense disc is misaligned by \sm40$^\circ$.  The gas is still infalling along the initial rotation axis, and is counter-rotating along the plane of the disc (i.e. $x \sim -y$).  These results are similar to that found in \citet{Tsukamoto+2017}.

Thus, there is a very small parameter space in which counter-rotating envelopes may form.  This required parameter space must include an initially slowly rotating envelope and strong magnetic fields that are either \bbmx, or  \bbmz \ if the Hall effect is included.

\subsection{Resolution}
\label{sec:disc:res}

Our models have been performed using a constant resolution of $10^6$ particles in the sphere, thus, we cannot explicitly discuss convergence, however, we will briefly comment on resolution.

Our mass resolution is $m_\text{p} = 10^{-6}$~\Msun \ per particle, thus there are \sm$10^5$ particles in our discs by the end of the simulation (recall Fig.~\ref{fig:disc:mass}).  Given this number of particles, the Jeans mass is still resolved \citep[][recall Section~\ref{sec:ic}]{BateBurkert1997}.

For discs, the Toomre-mass,
\begin{equation}
M_\text{T} = \frac{\pi c_\text{s}^4}{G^2 \Sigma},
\end{equation}
must be resolved to prevent numerically induced fragmentation.  From this equation, \citet{Nelson2006} calculated the maximum resolvable surface density to be 
\begin{equation}
\Sigma_\text{max} = \frac{\pi c_\text{s}^4}{G^2 m_\text{p} N_\text{reso}},
\end{equation}
which we have modified to 
\begin{equation}
\label{eq:sigmamax}
\Sigma_\text{m,max} = \frac{\pi \left(c_\text{s}^2+v_\text{A}^2\right)^2}{G^2 m_\text{p} N_\text{reso}}
\end{equation}
for our magnetised models.  Here, $N_\text{reso}$ is the number of particles required to resolve this maximum surface density, which  \citet{Nelson2006} empirically determined to be $N_\text{reso} \sim 6N_\text{neigh}$; given our cubic spline kernel, $N_\text{reso} \sim 342$ particles.  Throughout our suite, we find that the surface density of the discs is a few orders of magnitude lower than $\Sigma_\text{m,max}$, hence our discs meet the Toomre-mass criterion.

Resolving the vertical structure of discs is also important.  A poorly resolved vertical structure will underestimate the mid-plane density and hence gas pressure, which will inflate the discs and increase the likelihood of fragmentation.  For SPH simulations, \citet{Nelson2006} determined that at least four smoothing lengths $h$ are required per scale-height at the disc mid-plane.  Analogous to Section~\ref{sec:massratio}, we calculate the scale height using 
\begin{equation}
H = \frac{\sqrt{c_\text{s}^2+v_\text{A}^2}}{\Omega}.
\end{equation}

Using the mid-plane smoothing length, $H/h < 4$ for $r \lesssim 10$~au,  thus, the inner regions of the discs do not meet this criteria and may be under resolved.  This is to be expected given the presence of, and boundary effects caused by, the sink particle \citepeg{MachidaInutsukaMatsumoto2014,Wurster+2017}.  An under-resolved inner disc should not affect our general conclusions since the disc is not expected to fragment at such small radii.  For $r \gtrsim 10$~au,  $H/h > 4$ is typically satisfied, thus, we can be confident that our discs are vertically resolved.

\subsubsection{Convergence studies in the literature}
Although our discs are resolved (via the Jeans mass, Toomre-mass, and scale-height), resolution may still affect our results.  Convergence studies, especially of disc formation and fragmentation, have been performed frequently throughout the literature, and these studies have suggested that the fragmentation results are resolution-dependent \citepeg{MeruBate2011criteria,MeruBate2011converge,MeruBate2012,ForganPriceBonnell2017,Meyer+2018}.  While most studies have used parameterised cooling rates, \cite{Meyer+2018} performed radiation hydrodynamics simulations and also found that decreasing the resolution prevents fragmentation.  While increasing the resolution allows the disc to initially fragment at similar times in their fiducial and high resolution simulations, the future evolutions diverge, with more fragments forming at high resolution than fiducial resolution.

Independent of resolution, all of the hydrodynamic models in \citet{ForganPriceBonnell2017} fragmented.  For their magnetised models, increasing their particle number by a factor of two yielded an additional model that fragmented.  Resolution also played a role in the disc formation study of \citet{WursterPriceBate2016}: When modelling ideal MHD, smaller discs formed in higher resolution simulations, while in the models that included the non-ideal MHD processes, the disc masses differed by less than 5 per cent between their two resolutions.  

The above studies show that the convergence issue is persistent.  This issue arises in models starting from a pre-initialised disc and a molecular cloud core, studies investigating low-mass and high-mass star formation, and is independent of numerical method (i.e. SPH or a grid code).

Since our study meets the resolution criteria discussed above, models radiation hydrodynamics, and we do not model the evolution of the fragments or subsequent fragmentation, we believe that our qualitative results will be unaffected by resolutions.  Furthermore, non-ideal MHD models tend to be less sensitive to resolution than their ideal MHD counterparts.   However, for definitive quantitative results, a convergence study would be needed.

\subsection{Comparison to other studies}
\label{sec:disc:studies}

Most fragmentation studies, including ours, show that fragmentation does not occur at small disc radii \citepeg{StamatellosWhitworth2008,Boley2009,Clarke2009,ForganPriceBonnell2017}.  At low opacities, however, fragmentation is possible at small radii due to the higher cooling rate of dense gas and the shorter lifetime of the first hydrostatic code \citepeg{MeruBate2010,Bate2014,Bate2019}.

Magnetic fields are typically neglected in disc fragmentation studies, however, \citet{ForganPriceBonnell2017} initialise their discs with a toroidal magnetic field.  At the longest cooling time, no magnetised disc fragmented (although the hydrodynamics discs did), while at the shorter cooling times, all the magnetised discs fragmented.   Their discs were initialised to produce a considerable number of fragments (if unstable), rather than the few fragments that formed in the models we present here.  Nonetheless, they find that a large fraction of the hydrodynamical disc fragmented, whereas the magnetised discs fragmented in a narrow band.  In our models, there is no significant difference in the fragmentation distance between the hydrodynamic and magnetised models (see Fig.~\ref{fig:results:frag}).   Similar to \citet{ForganPriceBonnell2017}, we find the inclusion of magnetic fields stabilises the discs, and either delays or prevents fragmentation.  

One significant difference between the two studies is that their hydrodynamical and magnetised discs were initialised with the same parameters (e.g. radius, surface density profile) whereas our discs formed self-consistently, thus our magnetised discs were typically much smaller and less massive than the hydrodynamic discs, and this size difference in our models also likely contributed to the delayed or prevention of disc formation.

Although not explicitly investigating the fragmentation limit, \citet{Zhao+2018} investigated disc formation and fragmentation in the presence of magnetic fields, and included Ohmic resistivity and ambipolar diffusion.  Similar to here, they concluded that faster rotating discs promoted fragmentation, and that the disc was more likely to fragment in the presence of weak magnetic fields compared to strong fields.  They also find a diversity in where the fragments form and how they evolve. 

\section{Summary and conclusion}
\label{sec:conclusion}

We have presented a suite of simulations studying the formation and fragmentation of discs around protostars in the presence of magnetic fields.  Our models were initialised as 1M$_\odot$ Bonnor-Ebert spheres, which collapsed to form protostars typically surrounded by massive protostellar discs.  We followed the evolution until the final classification -- fragmented, spiral arms, smooth, or transient -- could be determined.  Our suite included ideal MHD, non-ideal MHD and purely hydrodynamical models, where the non-ideal MHD models included Ohmic resistivity, ambipolar diffusion and the Hall effect.  We tested five initial rotations $\Omega_0$, four initial magnetic field strengths $\mu_0$, and three (non-ideal MHD) or two (ideal MHD) orientations of the magnetic field.  Our simulations were radiation hydrodynamics simulations that were performed using the SPH code \textsc{sphng}.

Our key results are as follows:
\begin{enumerate}

\item Disc-like structures (herein referred to `discs') formed later for models with faster initial rotations and/or stronger magnetic fields.

\item Of our 105 models, 41 fragmented, 21 formed spiral structures but did not fragment, 12 formed smooth discs, and 31 did not form discs.   Discs were more likely to fragment for faster initial rotations (i.e. \omegage{0.45}), and for weaker magnetic fields (i.e. \muge{10}).  Non-ideal effects became important for strong magnetic fields (i.e. \mule{5}), and many of these discs with \bbpmz \ fragmented while their ideal MHD counterparts did not.  

\item For the discs that fragmented, there was no clear correlation between fragmentation time or distance and our initial parameters (rotation rate, magnetic field strength and orientation). 

\item The magnetic Toomre-Q parameter, $Q_\text{m}$, cannot be used in isolation in our models to determine if a model will fragment.  All models that fragment do so in a region of $Q_\text{m} < 1.4$, but not all models that entered this regime fragmented.  When comparing the ratio of disc-to-stellar masses, the ratio suggested that all of our discs were unstable to fragmentation.  This is a result of the young protostar that is still accreting mass from the disc.  

\item Outflows were launched from models with initially strong, \bbpmz \ magnetic fields.  Fewer models that include the non-ideal MHD processes launch outflows since these processes weaken the magnetic field.

\item Counter-rotating envelopes form only under specific conditions: an initially slowly rotating envelope with strong magnetic fields that are either \bbmx, or \bbmz \ if the Hall effect is included.

\item Discs masses up to \sm0.1~\Msun were obtained (i.e., up to \sm10~per cent of the SPH particles that were initially in the sphere).  The Jeans mass and Toomre mass were resolved throughout the calculations, and the scale-height was resolved for $r > 10$~au.  However, a proper resolution study would be required for a thorough discussion of convergence.

\end{enumerate}

Given our results, we cannot predict when or where a disc will fragment.  However, qualitatively, fragmentation is promoted in faster rotating models and in models with weaker magnetic fields.

\section*{Acknowledgements}

We would like to thank the referee for useful and insightful comments that improved the quality of this manuscript.
JW and MRB acknowledge support from the European Research Council under the European Community's Seventh Framework Programme (FP7/2007- 2013 grant agreement no. 339248).  The calculations for this paper were performed on the University of Exeter Supercomputers, Zen and Isca.  The former was a DiRAC Facility jointly funded by STFC, the Large Facilities Capital Fund of BIS, and the University of Exeter; the latter is part of the University of Exeter High-Performance Computing (HPC) facility.  Analyses were performed the DiRAC Complexity system, operated by the University of Leicester IT Services, which forms part of the STFC DiRAC HPC Facility (www.dirac.ac.uk). This equipment is funded by BIS National E-Infrastructure capital grant ST/K000373/1 and  STFC DiRAC Operations grant ST/K0003259/1. Additional analyses were performed using using the DiRAC Data Intensive service at Leicester, operated by the University of Leicester IT Services, which forms part of the STFC DiRAC HPC Facility (www.dirac.ac.uk). The equipment was funded by BEIS capital funding via STFC capital grants ST/K000373/1 and ST/R002363/1 and STFC DiRAC Operations grant ST/R001014/1. DiRAC is part of the National e-Infrastructure.
The column density figures were made using {\sc splash} \citep{Price2007}.
The research data supporting this publication are openly available from the University of Exeter's institutional repository, https://ore.exeter.ac.uk/repository.

\appendix
\section{Appendix}
\label{app:results}
Tables \ref{table:app:results:HI} and \ref{table:app:results:N} summarise the models at the end of the simulation.  The simulations are ended once the final classification can be determined.    A few models have small, transient asymmetries; although these features are not well-defined and persistent spiral arms, it means the disc is not completely smooth.  For simplicity, we give these models the classification of the most persistent state, and identify them with an asterisk after their classification in the following tables.  Along with its classification, each model is listed with its disc formation time, the end-time relative to the disc formation time, and the radius and mass of both the bulk and total discs at this time.  The disc properties are calculated as described in Section~\ref{sec:results:identify:disc}; recall that `disc' refers to `disc-like structure.'  For the models that fragment, the data we present is from the first output that contains the fragment; for the transient models, the data we present is from the first output in which a disc is not present.  If a disc never forms, we present the sink formation time, and the relative end-time is set to zero.  Also included is the stellar mass (i.e the mass of the sink particle).   The penultimate column indicates if a fast  ($v_\text{r} \sim 1-5$~\kms) or slow ($v_\text{r} < 1$~\kms) outflow existed at any time during the simulation, and the final column indicates if a counter-rotating envelope (CRE) existed at any time during the simulation. 

\begin{table*}
\begin{center}
\begin{tabular}{l l c c c c c c c c l l}
\hline
Name & Classification & $t_\text{disc}$ & $t_\text{end}-t_\text{disc}$ & $d_\text{frag}$ & $R_\text{B,disc}$ & $R_\text{T,disc}$ & $M_\text{B,disc}$ & $M_\text{T,disc}$ & $M_\text{star}$ & Outflow & CRE \\
          &                 &     [kyr]    & [kyr]  & [au] & [au]  & [au]  & [\Msun] & [\Msun] & [\Msun] &   &   \\
\hline
\hmodel{5}   & Transient & 116 & 0     & -     & - & - & -  & - & - & no & no \\ 
\hmodel{25} & Spiral Arms & 119 &   17.2  & -     & 54.0 & 75.9 & 0.138  & 0.152 & 0.147 & no & no \\
\hmodel{45} & Fragmented & 125 &  15.7 & 137 & 47.0 &  149 & 0.0836  & 0.112 & 0.0854 & no & no \\
\hmodel{65} & Fragmented & 136 &  2.51 & 50.0 & 14.5 & 67.8 & 0.0356  & 0.0496 & 0.00208 & no & no \\
\hmodel{85} & Fragmented & 154 &  3.22 & 83.9 & 17.5 & 92.4 & 0.0329  & 0.0463 & 0.00362 & no & no \\
\\ %
\imodel{5}{3}{-z} & Transient          & 123 &  0    & -        &    -       &    -         & -            & -           & - & no & no \\ 
\imodel{5}{5}{-z} & Transient          & 119 & 0.716      & -       &    -       &    -         & -            & -           & 0.0273 & no & no \\ 
\imodel{5}{10}{-z} & Transient          & 117 & 0.358    & -        &    -       &    -         & -            & -           & 0.0226 & no & no \\ 
\imodel{5}{20}{-z} & Transient   & 117 & 0.358  & -        &- & - & -  & - & 0.0251 & no & no \\ 
\imodel{25}{3}{-z} & Transient  & 126 & 0  & -        &    -     &   - &- & -& - & slow & no \\  
\imodel{25}{5}{-z} & Transient  & 121 & 5.37  & -        &- & - & -  & - & 0.112 & slow & no \\
\imodel{25}{10}{-z} & Smooth & 120 &15.8 & -        &42.0 & 60.7 & 0.113  & 0.122 & 0.151 & no & no \\ 
\imodel{25}{20}{-z} & Spiral Arms & 119 & 16.1  & -        &40.0 & 77.3 & 0.104  & 0.120 & 0.153 & no & no \\
\imodel{45}{3}{-z} & Transient   & 131 &3.94 & -     &  -       &    -      &    -   & -& 0.0816 & slow & no \\  
\imodel{45}{5}{-z} & Smooth & 127 &15.8  & -        &32.0 & 57.4 & 0.0853  & 0.0949 & 0.118 & slow & no \\
\imodel{45}{10}{-z} & Fragmented & 125  &   5.50 & 51.0 & 35.0 & 98.1 & 0.0498  & 0.0714 & 0.0484 & no & no \\
\imodel{45}{20}{-z} & Fragmented & 125 &   8.80 & 87.3 & 43.5 & 129 & 0.0693  & 0.0877 & 0.0632 & no & no \\
\imodel{65}{3}{-z} & Transient  & 143 & 4.30  &-      & -        &    -       &    -         & - & 0.0815 & slow & no \\
\imodel{65}{5}{-z} & Spiral Arms & 139 & 15.8 & -        &36.5 & 70.4 & 0.0516  & 0.0663 & 0.0932 & slow & no \\
\imodel{65}{10}{-z} & Fragmented & 136 &   2.86 & 57.0 & 28.0 & 76.6 & 0.0234  & 0.0406 & 0.0244 & no & no \\
\imodel{65}{20}{-z} & Fragmented & 136 &   3.58 & 88.0 & 33.0 & 102 & 0.0300  & 0.0557 & 0.0280 & no & no \\
\imodel{85}{3}{-z} & Smooth   & 162 & 15.8  & -        &    12.0       &    15.4         & 0.00356           & 0.00388 & 0.137 & slow & no \\  
\imodel{85}{5}{-z} & Fragmented & 158   &  3.22 & 34.0 & 35.5 & 51.3 & 0.0369  & 0.0408 & 0.0236 & slow & no \\
\imodel{85}{10}{-z} & Fragmented & 156  &   1.79 & 58.2 & 25.5 & 75.6 & 0.0178  & 0.037 & 0.0172 & no & no \\
\imodel{85}{20}{-z} & Fragmented & 155  &  2.51 & 72.8 & 24.5 & 91.7 & 0.0181  & 0.0355 & 0.0162 & no & no \\
\\ %
\imodel{5}{3}{-x} & Transient          & 122 & 0.716  & -        &    -       &    -         & -            & -           & 0.0404 & no & yes \\ 
\imodel{5}{5}{-x} & Transient          & 119 &  0.358 & -        &    -       &    -         & -            & -           & 0.0264 & no & no \\ 
\imodel{5}{10}{-x} & Transient          & 117 & 0.358 & -       &    -       &    -         & -            & -           & 0.0236 & no & no \\ 
\imodel{5}{20}{-x} & Transient          & 117 &  0.358 & -       &    -       &    -         & -            & -           & 0.0238 & no & no \\ 
\imodel{25}{3}{-x} & Transient          & 125 & 0.358 & -        &    -       &    -         & -            & -           & 0.0240 & no & no \\ 
\imodel{25}{5}{-x} & Transient          & 121 & 1.07  & -        &- & - & -  & - & 0.0419 & no & no \\
\imodel{25}{10}{-x} & Smooth & 120 &15.8 & -        &32.5  & 43.0 & 0.101 & 0.105 & 0.193 & no & no \\
\imodel{25}{20}{-x} & Spiral Arms & 119 & 15.8  & -        &43.5 & 79.7 & 0.103  & 0.116 & 0.165 & no & no \\
\imodel{45}{3}{-x} & Transient          & 132 & 0.716 & -  &    -       &    -         & -            & -           & 0.0274 & no & no \\
\imodel{45}{5}{-x} & Smooth & 128 & 15.8 & -        &34.5 & 41.0 & 0.0915  & 0.0938 & 0.168 & no & no \\
\imodel{45}{10}{-x} & Fragmented & 126  &   12.1 & 150 & 38.5 & 158 & 0.0848  & 0.0983 & 0.107 & no & no \\
\imodel{45}{20}{-x} & Fragmented & 125  &   11.0 & 70.0 & 39.5 & 200 & 0.0695  & 0.0899 & 0.0797 & no & no \\
\imodel{65}{3}{-x} & Transient          & 145 & 0.716  & - &    -  &    -         & -            & -           & 0.0237 & no & no \\
\imodel{65}{5}{-x} & Spiral Arms & 140 & 15.8  & - & 39.5 & 65.4 & 0.0737  & 0.0881 & 0.127 & no & no \\
\imodel{65}{10}{-x} & Fragmented & 137 &  11.8 & 79.6 & 41.0 & 110 & 0.0625  & 0.0868 & 0.0783 & no & no \\
\imodel{65}{20}{-x} & Fragmented & 136 &  6.80 & 36.0 & 38.5 & 111 & 0.0522  & 0.0578 & 0.0433 & no & no \\
\imodel{85}{3}{-x} & Transient          & 168 &  0.358   & -        &    - &    -  & -            & -           & 0.0224 & no & no \\
\imodel{85}{5}{-x} & Spiral Arms & 162 & 15.8         & -        &44.5 & 70.0 & 0.0685  & 0.0788 & 0.114 & no & no \\
\imodel{85}{10}{-x} & Fragmented & 158 &   14.3 & 71.0 & 45.0 & 109 & 0.0579  & 0.0729 & 0.0729 & no & no \\
\imodel{85}{20}{-x} & Fragmented & 155 &   4.30 & 55.1 & 34.5 & 67.1 & 0.0314  & 0.0449 & 0.0259 & no & no \\
\hline
\end{tabular}
\caption{Summary of the results at the end of the hydrodynamics and ideal MHD models.  The columns are as defined throughout the text and summarised at the beginning of this appendix.} 
\label{table:app:results:HI} 
\end{center}
\end{table*}
\begin{table*}
\begin{center}
\begin{tabular}{l l c c c c c c c c l l}
\hline
Name & Classification & $t_\text{disc}$  &$t_\text{end}-t_\text{disc}$ & $d_\text{frag}$ & $R_\text{B,disc}$ & $R_\text{T,disc}$ & $M_\text{B,disc}$ & $M_\text{T,disc}$ & $M_\text{star}$ & Outflow & CRE \\
          &                 &     [kyr]    & [kyr]  & [au] & [au]  & [au]  & [\Msun] & [\Msun] & [\Msun] &   &   \\
\hline
\nmodel{5}{3}{-z} & Transient & 122 &0.716   & -        &- & - & -  & - & 0.0307 & no & yes \\  
\nmodel{5}{5}{-z} & Transient & 119 & 0 & -        &- & - & -  & - & - & no & yes \\
\nmodel{5}{10}{-z} & Transient & 117 &0.358   & -        &- & - & -  & - & 0.0237 & no & no \\ 
\nmodel{5}{20}{-z} & Transient & 116 &0.716  & -        &- & - & -  & - & 0.0254 & no & no \\   
\nmodel{25}{3}{-z} & Smooth & 125 &15.4   & -        &32.0 & 39.4 & 0.107  & 0.108 & 0.190 & slow & no \\
\nmodel{25}{5}{-z} & Spiral Arms & 122 & 15.8& -        &41.0 & 75.8 & 0.104  & 0.117 & 0.173 & no & no \\
\nmodel{25}{10}{-z} & Spiral Arms & 120&15.8 & -        &50.5 & 71.2 & 0.132  & 0.148 & 0.154 & no & no \\
\nmodel{25}{20}{-z} & Spiral Arms & 119 & 15.8  & -        &41.5 & 79.4 & 0.0974  & 0.124 & 0.145 & no & no \\
\nmodel{45}{3}{-z} & Smooth* & 131 &  15.7  & -        & 37.0 & 57.0 & 0.0923  & 0.0994 & 0.145 & fast & no \\ 
\nmodel{45}{5}{-z} & Fragmented & 127  & 10.2 & 73.8 & 46.0 & 113 & 0.0864  & 0.103 & 0.0777 & no & no \\
\nmodel{45}{10}{-z} & Fragmented & 125 & 8.25 & 85.2 & 36.5 & 111 & 0.0547  & 0.0861 & 0.0581 & no & no \\
\nmodel{45}{20}{-z} & Fragmented & 125 & 8.80 & 41.2 & 45.0 & 103 & 0.0713  & 0.102 & 0.0587 & no & no \\
\nmodel{65}{3}{-z} & Spiral Arms & 142 & 14.9 & -        &33.5 & 53.1 & 0.0788  & 0.0850 & 0.115 & fast & no \\ 
\nmodel{65}{5}{-z} & Fragmented & 138  & 3.27 & 57.0 & 25.0 & 96.2 & 0.0354  & 0.0525 & 0.0236 & no & no \\
\nmodel{65}{10}{-z} & Fragmented & 136 & 3.16 & 76.4 & 28.5 & 94.8 & 0.0522  & 0.0692 & 0.0355 & no & no \\
\nmodel{65}{20}{-z} & Fragmented & 135 & 2.80 & 49.5& 14.5 & 62.7 & 0.0357  & 0.0569 & 2.68$\times10^{-4}$ & no & no \\
\nmodel{85}{3}{-z} & Fragmented & 162  & 3.07 & 64.9& 27.0 & 74.7 & 0.0353  & 0.0466 & 0.0207 & slow & no \\
\nmodel{85}{5}{-z} & Fragmented & 168   & 2.96   & 68.7 & 23.5 & 79.5 & 0.0446  & 0.0628 & 3.32$\times10^{-3}$ & no & no \\
\nmodel{85}{10}{-z} & Fragmented & 155 & 2.86 & 48.8 & 23.5 & 84.8 & 0.0381  & 0.0564 & 2.04$\times10^{-3}$ & no & no \\
\nmodel{85}{20}{-z} & Fragmented & 154 & 3.16 & 52.3 & 25.5 & 91.9 & 0.0349  & 0.0478 & 2.97$\times10^{-3}$ & no & no \\
\\ %
\nmodel{5}{3}{+z} & Transient          & 122 & 0.716   & -        &    -       &    -         & -            & -           & 0.0293 & fast & no \\ 
\nmodel{5}{5}{+z} & Transient          & 119 & 0.716  & -        &    -       &    -         & -            & -           & 0.0291 & no & no \\ 
\nmodel{5}{10}{+z} & Transient          & 117 & 0.358  & -        &    -       &    -         & -            & -           & 0.0231 & no & no \\
\nmodel{5}{20}{+z} & Transient         & 116 &0.716  & -        &    -       &    -         & -            & -          & 0.0261 & no & no \\ 
\nmodel{25}{3}{+z} & Transient          & 125 & 0   & -        &    -       &    -         & -            & -           & 0 & fast & no \\  
\nmodel{25}{5}{+z} & Smooth* & 122 & 15.4   & -        &39.0 & 57.7 & 0.107  & 0.118 & 0.178 & no & no \\ 
\nmodel{25}{10}{+z} & Fragmented & 120 &  11.8 & 83.2 & 38.5 & 108 & 0.100  & 0.111 & 0.127 & no & no \\
\nmodel{25}{20}{+z} & Spiral Arms & 119& 15.8 & -        &42.5 & 81.4 & 0.107  & 0.125 & 0.142 & no & no \\
\nmodel{45}{3}{+z} & Smooth & 131 & 15.3  & -        &29.0 & 31.8 & 0.0776 & 0.0780 & 0.174 & fast & no \\
\nmodel{45}{5}{+z} & Fragmented & 127   & 7.94 & 69.2 & 36.0 & 132 & 0.0628  & 0.0841 & 0.0679 & no & no \\
\nmodel{45}{10}{+z} & Fragmented & 125  & 6.09 & 108 & 35.0 & 118 & 0.0535  & 0.0721 & 0.0445 & no & no \\
\nmodel{45}{20}{+z} & Fragmented & 125  & 8.97& 71.9 & 36.0 & 104 & 0.0556  & 0.0896 & 0.0615 & no & no \\
\nmodel{65}{3}{+z} & Smooth* & 142 & 15.2  & -        & 31.5 & 37.9 & 0.0878 & 0.0892 & 0.132 & fast & no \\ 
\nmodel{65}{5}{+z} & Fragmented & 138  & 2.86 & 72.4 & 25.5 & 81.1 & 0.0456  & 0.0601 & 8.52$\times10^{-3}$ & no & no \\
\nmodel{65}{10}{+z} & Fragmented & 136  & 2.73 & 55.0 & 15.0 & 70.9 & 0.0361  & 0.0568 & 7.78$\times10^{-4}$ & no & no \\
\nmodel{65}{20}{+z} & Fragmented & 135  & 2.84 & 47.4 & 15.0 & 73.4 & 0.0362  & 0.0583 & 1.82$\times10^{-4}$ & no & no \\
\nmodel{85}{3}{+z} & Fragmented & 162    & 3.58 & 51.8 & 26.0 & 81.8 & 0.0355  & 0.0456 &          0.0259           & slow & no \\
\nmodel{85}{5}{+z} & Fragmented & 158   & 2.73 & 62.6 & 21.5 & 77.2 & 0.0410  & 0.0607 & 1.60$\times10^{-3}$ & no & no \\
\nmodel{85}{10}{+z} & Fragmented & 155 & 2.80 & 37.3 & 15.5 & 74.0 & 0.0311  & 0.0511 & 2.75$\times10^{-4}$ & no & no \\
\nmodel{85}{20}{+z} & Fragmented & 154   & 3.49 & 55.7 & 25.5 & 84.4 & 0.0372  & 0.0589 & 3.08$\times10^{-3}$ & no & no \\
\\ %
\nmodel{5}{3}{-x} & Transient          & 122 & 0.358  & -        &    -       &    -         & -            & -           & 0.0214 & no & yes \\
\nmodel{5}{5}{-x} & Transient          & 119 & 0.716     & -        &- & - & -  & - & 0.0287 & no & no \\  
\nmodel{5}{10}{-x} & Transient  & 117 & 0.358  & -        &- & - & -  & - & 0.0217 & no & no \\ 
\nmodel{5}{20}{-x} & Transient & 117 &0.358  & -        &- & - & -  & - & 0.0241 & no & no \\
\nmodel{25}{3}{-x} & Smooth & 126 & 15.8   & -        &31.5 & 36.5 & 0.102  & 0.103 & 0.211 & no & no \\
\nmodel{25}{5}{-x} & Smooth & 122 & 15.8  & -        & 42.0 & 52.2 & 0.124  & 0.128 & 0.176 & no & no \\
\nmodel{25}{10}{-x} & Spiral Arms & 120 & 15.8 & -        &50.0 & 83.4 & 0.112  & 0.132 & 0.162 & no & no \\
\nmodel{25}{20}{-x} & Spiral Arms & 119 & 15.9 & -        &47.0 & 98.4 & 0.112  & 0.138 & 0.144 & no & no \\
\nmodel{45}{3}{-x} & Spiral Arms & 132  & 16.0  & -        &46.0 & 59.5 & 0.108  & 0.117 & 0.160 & no & no \\ 
\nmodel{45}{5}{-x} & Spiral Arms & 128  & 15.8  & -        & 41.0 & 84.5 & 0.0796  & 0.113 & 0.126& no & no \\ 
\nmodel{45}{10}{-x} & Spiral Arms & 125 &15.6 & -        &41.0 & 83.8 & 0.0803  & 0.101 & 0.0982 & no & no \\
\nmodel{45}{20}{-x} & Spiral Arms & 125 & 16.0 & -        &48.5 & 112 & 0.0867  & 0.118 & 0.0887 & no & no \\
\nmodel{65}{3}{-x} & Spiral Arms & 145 & 15.4  & -        &39.5 & 67.9 & 0.0749  & 0.102 & 0.134 & no & no \\  
\nmodel{65}{5}{-x} & Spiral Arms & 140 & 15.6  & -        &41.5 & 96.8 & 0.0651  & 0.0954 & 0.0982 & no & no \\ 
\nmodel{65}{10}{-x} & Fragmented & 136  & 11.9 & 105 & 40.0 & 143 & 0.0605  & 0.0822 & 0.0611 & no & no \\
\nmodel{65}{20}{-x} & Fragmented & 135  & 3.44 & 41.3 & 28.0 & 75.3 & 0.0443  & 0.0542 & 0.0899 & no & no \\
\nmodel{85}{3}{-x} & Spiral Arms & 168  & 15.5 & -        &53.0& 62.9 & 0.102  & 0.110 & 0.131 & no & no \\ 
\nmodel{85}{5}{-x} & Spiral Arms & 162 & 15.4  & -        & 42.5 & 99.3 & 0.0545  & 0.0855 & 0.0900 & no & no \\
\nmodel{85}{10}{-x} & Fragmented & 157  & 11.6 & 93.0 & 34.5 & 129 & 0.0325  & 0.0646 & 0.0526 & no & no \\
\nmodel{85}{20}{-x} & Fragmented & 155  & 3.47 & 32.5 & 23.5 & 79.6 & 0.0396  & 0.0534 & 0.0591 & no & no \\
\hline
\end{tabular}
\caption{Summary of the results at the end of the non-ideal MHD models.  The columns are as defined throughout the text and summarised at the beginning of this appendix.} 
\label{table:app:results:N} 
\end{center}
\end{table*}

\bibliography{DiscFormFrag.bib}

\label{lastpage}
\end{document}